\documentclass[prl,superscriptaddress,a4paper,twocolumn]{revtex4-1}
\usepackage{geometry}
\usepackage[english]{babel}
\usepackage[latin1]{inputenc}
\usepackage[colorlinks,linkcolor=blue,anchorcolor=blue,citecolor=blue]{hyperref}
\usepackage{amsfonts}
\usepackage{amsmath, amssymb, amsfonts, mathrsfs}
\usepackage{graphicx}
\usepackage{xspace}
\usepackage{xcolor}
\usepackage{multirow}

\usepackage{array} 
\newcolumntype{P}[1]{>{\centering\arraybackslash}p{#1}}
\newcolumntype{M}[1]{>{\centering\arraybackslash}m{#1}}

\usepackage[bottom]{footmisc}

\newcommand{\ket}[1]{|#1\rangle}
\newcommand{\bra}[1]{ \langle #1 \,|}

\begin{document}

\title{Heralded non-destructive quantum entangling gate with single-photon sources}

\author{Jin-Peng Li} 
\author{Xuemei Gu}
\author{Jian Qin}
\author{Dian Wu}
\author{Xiang You}
\author{Hui Wang}
\affiliation{Hefei National Laboratory for Physical Sciences at Microscale and Department of Modern Physics, University of Science and Technology of China, Hefei, Anhui 230026, China}
\affiliation{CAS Center for Excellence and Synergetic Innovation Center in Quantum Information and Quantum Physics, University of Science and Technology of China, Shanghai 201315, China}

\author{Christian Schneider}
\affiliation{Institute of Physics, Carl von Ossietzky University, 26129 Oldenburg, Germany} 
\affiliation{Technische Physik, Physikalische Institut and Wilhelm Conrad R{\"{o}}ntgen-Center for Complex Material Systems, Universit{\"{a}}t W{\"{u}}rzburg, Am Hubland, D-97074 W{\"{u}}rzburg, Germany}

\author{Sven H{\"{o}}fling}
\affiliation{CAS Center for Excellence and Synergetic Innovation Center in Quantum Information and Quantum Physics, University of Science and Technology of China, Shanghai 201315, China}
\affiliation{Technische Physik, Physikalische Institut and Wilhelm Conrad R{\"{o}}ntgen-Center for Complex Material Systems, Universit{\"{a}}t W{\"{u}}rzburg, Am Hubland, D-97074 W{\"{u}}rzburg, Germany}

\author{Yong-Heng Huo}
\affiliation{Hefei National Laboratory for Physical Sciences at Microscale and Department of Modern Physics, University of Science and Technology of China, Hefei, Anhui 230026, China}
\affiliation{CAS Center for Excellence and Synergetic Innovation Center in Quantum Information and Quantum Physics, University of Science and Technology of China, Shanghai 201315, China}

\author{Chao-Yang Lu}
\affiliation{Hefei National Laboratory for Physical Sciences at Microscale and Department of Modern Physics, University of Science and Technology of China, Hefei, Anhui 230026, China}
\affiliation{CAS Center for Excellence and Synergetic Innovation Center in Quantum Information and Quantum Physics, University of Science and Technology of China, Shanghai 201315, China}

\author{Nai-Le Liu}
\affiliation{Hefei National Laboratory for Physical Sciences at Microscale and Department of Modern Physics, University of Science and Technology of China, Hefei, Anhui 230026, China}
\affiliation{CAS Center for Excellence and Synergetic Innovation Center in Quantum Information and Quantum Physics, University of Science and Technology of China, Shanghai 201315, China}

\author{Li Li}
\email{eidos@ustc.edu.cn}
\affiliation{Hefei National Laboratory for Physical Sciences at Microscale and Department of Modern Physics, University of Science and Technology of China, Hefei, Anhui 230026, China}
\affiliation{CAS Center for Excellence and Synergetic Innovation Center in Quantum Information and Quantum Physics, University of Science and Technology of China, Shanghai 201315, China}

\author{Jian-Wei Pan}
\affiliation{Hefei National Laboratory for Physical Sciences at Microscale and Department of Modern Physics, University of Science and Technology of China, Hefei, Anhui 230026, China}
\affiliation{CAS Center for Excellence and Synergetic Innovation Center in Quantum Information and Quantum Physics, University of Science and Technology of China, Shanghai 201315, China}

\begin{abstract}
Heralded entangling quantum gates are an essential element for the implementation of large-scale optical quantum computation. Yet, the experimental demonstration of genuine heralded entangling gates with free-flying output photons in linear optical system, was hindered by the intrinsically probabilistic source and double-pair emission in parametric down-conversion. Here, by using an on-demand single-photon source based on a semiconductor quantum dot embedded in a micro-pillar cavity, we demonstrate a heralded controlled-NOT (CNOT) operation between two single photons for the first time. To characterize the performance of the CNOT gate, we estimate its average quantum gate fidelity of ($87.8\pm1.2$)\%. As an application, we generated event-ready Bell states with a fidelity of ($83.4\pm2.4$)\%. Our results are an important step towards the development of photon-photon quantum logic gates. 
\end{abstract}
\date{\today}
\maketitle
 
Entangling gates are a crucial building block in scalable quantum computation, as it enables the construction of any quantum computing circuits when combined with single-qubit gates \cite{nielsen2010quantum}. One of the canonical example is the controlled-NOT (CNOT) gate, which flips the target qubit state conditional on the control qubit. Photons are generally accepted as the best candidate for a qubit due to their negligible decoherence and ease of single-qubit operation. Unfortunately, ambitions to implement optical CNOT gates are hampered as they require strong interactions between individual photons well beyond those presently available. Surprisingly, projective measurements with photodetector can induce an effective nonlinearity sufficient for the realization of entangling gates using linear optics \cite{knill2001scheme}. Since then, many schemes to implement optical CNOT gates have been theoretically proposed \cite{ralph2001simple, ralph2002linear, pittman2001probabilistic} and experimentally demonstrated \cite{o2003demonstration, pittman2002demonstration, pittman2003experimental,langford2005demonstration, okamoto2005demonstration, okamoto2011realization, gasparoni2004realization, zhao2005experimental, huang2004experimental, gao2010teleportation, zeuner2018integrated,bao2007optical}.

Early demonstrations came at the expense of destroying the output states by detecting the photons\cite{o2003demonstration, pittman2002demonstration, pittman2003experimental,langford2005demonstration, okamoto2005demonstration, okamoto2011realization}, thus limiting the scaling to a larger system. To be scalable, heralded CNOT gates are necessary \cite{knill2001scheme,gasparoni2004realization}. Specifically, a successful operation is heralded by the detection of ancillary photons. Such heralded gates are highly important as they provide information classically feed forwardable, which is crucial for a scalable architecture both in the standard circuit model \cite{nielsen2010quantum, knill2001scheme} and one-way model using cluster states \cite{raussendorf2001one, walther2005experimental, rudolph2017optimistic}. Implementations of heralded CNOT gates assisted with entangled \cite{gasparoni2004realization, zhao2005experimental, huang2004experimental, gao2010teleportation, zeuner2018integrated} or single ancilla photons \cite{bao2007optical} have been reported. However, the demonstrations mainly employed probabilistic photon sources such as pair sources based on spontaneous parametric down-conversion (SPDC) \cite{burnham1970observation}. Due to the probabilistic nature of SPDC that involves multiphoton emissions, one cannot obtain a photon pair or a heralded single photon deterministically with high generation rate \cite{couteau2018spontaneous}, and daunting scalability issues raise if we wish to use these photons. Also, as multiple pair events are always present and detectors without photon number resolution are commonly used, which introduce false heralding signals, post-selection is necessary in the demonstrations to confirm a successful gate operation \cite{pan2003experimental}. For this reason, while the schemes in principle work in a heralded way, the previous experiments are actually a destructive version of the heralded CNOT gates, limiting their further applications. To overcome these issues, on-demand photon sources will be the fundamental assets.

Semiconductor quantum dots (QDs) confined in a micro-cavity are particularly appealing emitters of on-demand photons, which can deterministically emit single photons \cite{lodahl2015interfacing, unsleber2016highly, somaschi2016near, ding2016demand} as well as entangled photon pairs \cite{wang2019demand, liu2019solid}. They have been shown to generate single photons simultaneously exhibiting high brightness, near-unity single-photon purity and indistinguishability \cite{ding2016demand, wang2019towards}, and currently have the best all-around single-photon source-performance \cite{senellart2017high, tomm2020bright}. Until now, QD single-photon sources (SPSs) have already been used to realize optical CNOT gates in a semiconductor waveguide chip \cite{pooley2012controlled} and bulk optics with partial polarizing beam splitters (PBSs) \cite{he2013demand} and multi-path interference \cite{gazzano2013entangling}. Nevertheless, none of them are heralded since they necessarily destroy the output states.

In this Letter, we demonstrate the first heralded CNOT operation between two single photons using a QD coupled in a micro-pillar cavity. Under pulse resonant excitation, the QD SPS generates high-quality single photons that are deterministically demultiplexed into four indistinguishable SPSs (two serve as input photons and two are ancillary photons) for CNOT gates. Heralded by the detection of two ancillary photons, we achieved a heralded CNOT gate with a fidelity of $(87.8\pm1.2)$\% and $\sim$85 operations per minute increased by at least an order of magnitude compared to early experiments, and generated an event-ready Bell state with a fidelity of $(83.4\pm2.4)\%$. Our results are an important step towards scalable optical quantum information processing (QIP). 

\begin{figure}[!t]
	\includegraphics[width=0.38\textwidth]{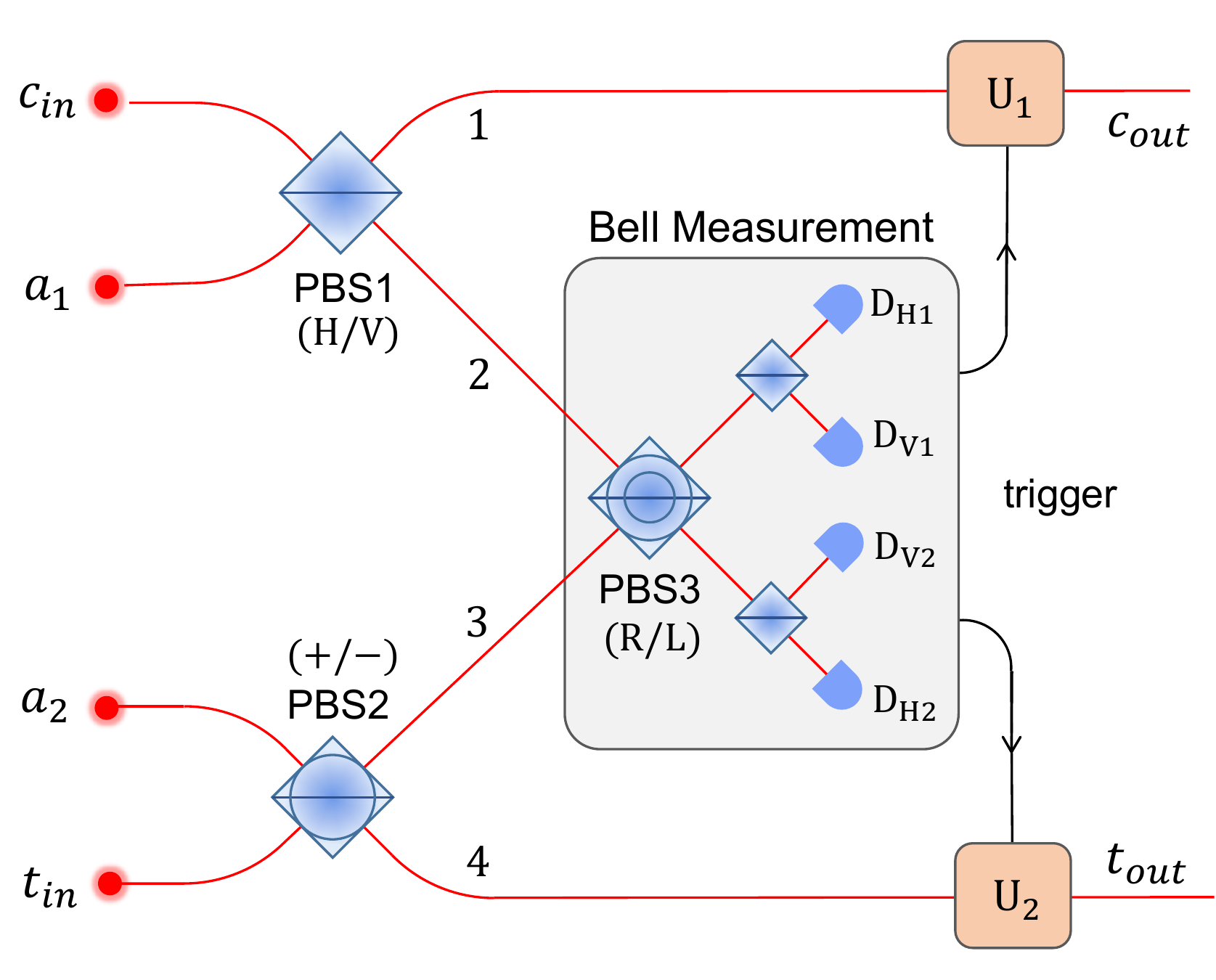}
	\caption{An optical heralded CNOT gate \cite{bao2007optical}. Polarizing beam splitters (PBSs) are used, where PBS1 (PBS2, PBS3) transmit $\ket{H}$ ($\ket{+}$, $\ket{R}$) states and reflect $\ket{V}$ ($\ket{-}$, $\ket{L}$) ones. The heralded CNOT gate depends on the ancilla measurement outcome and related Pauli operators $U$.}  
	\label{fig:Concept}
\end{figure}

The scheme for a heralded CNOT gate requires only four single photons and three polarizing beam splitters (PBSs) in mutually unbiased bases as described in Fig.\ref{fig:Concept}. We prepare polarization-encoded photons, and define horizontal $\ket{H}$ and vertical $\ket{V}$ polarization as logical states $\ket{0}$ and $\ket{1}$. The input state of the control and target qubits are $\ket{\psi}_{c_{in}}=\alpha\ket{H}+\beta\ket{V}$ and $\ket{\psi}_{t_{in}}=\gamma\ket{H}+\delta\ket{V}$, where complex coefficients $\alpha$ and $\beta$ ($\gamma$ and $\delta$) satisfy $|\alpha|^2+|\beta|^2=1$ ($|\gamma|^2+|\delta|^2=1$), and subscript represents the photon's path (holding for the rest of the paper). Two ancillary single photons are in the states of $\ket{\psi}_{a_{1}}=1/\sqrt{2}\left(\ket{H}+\ket{V}\right)$ and $\ket{\psi}_{a_{2}}=\ket{H}$\cite{bao2007optical,supp}. The four single photons are then superimposed at PBS1 and PBS2, as shown in Fig.\ref{fig:Concept}. When there is only one photon in each path after PBS1 and PBS2, we obtained a four-photon state with a probability of 1/4, which is further rewritten as follows (see the supplemental material for details\cite{supp}):
\begin{align}
\ket{\Psi}= \nonumber&(I_{1}I_{4}U_{14}\ket{\psi}_{1}^{c_{in}}\ket{\psi}_{4}^{t_{in}}\otimes\ket{\Phi^{+}}_{23}\\ \nonumber
& +I_{1}\sigma_{x4}U_{14}\ket{\psi}_{1}^{c_{in}}\ket{\psi}_{4}^{t_{in}}\otimes\ket{\Psi^{+}}_{23}\\ \nonumber
&+ \sigma_{z1}I_{4}U_{14}\ket{\psi}_{1}^{c_{in}}\ket{\psi}_{4}^{t_{in}}\otimes\ket{\Phi^{-}}_{23}\\ 
& +\sigma_{z1}\sigma_{x4}U_{14}\ket{\psi}_{1}^{c_{in}}\ket{\psi}_{4}^{t_{in}}\otimes\ket{\Psi^{-}}_{23})/2
\label{eq:CNOTstate}
\end{align}
where $U_{14}$ refers to the CNOT operation between photons in path $c_{in}$ and $t_{in}$ (further refer to path 1 and 4), and the output photons in path $c_{out}$ and $t_{out}$ describe the result of CNOT gates; $\ket{\Phi^{\pm}}$ and $\ket{\Psi^{\pm}}$ are standard Bell states in $\ket{H}/\ket{V}$ basis; $I_{i}$, $\sigma_{zi}$ and $\sigma_{xi}$ are Identity, Pauli Z and Pauli X operations on the photon in path $i$ ($i=1,4$).

Obviously, one can get the CNOT operation $U_{14}$ by performing a jointly projective measurement of Bell states on ancillary photons in path 2 and 3 together with their related Pauli operations on the output photons in path 1 and 4, as described in Eq.(\ref{eq:CNOTstate}). For standard linear optical Bell-state analyser, only two of the four Bell states can be distinguished \cite{calsamiglia2001maximum}. In Fig.\ref{fig:Concept}, two Bell states are $\ket{\Psi^{+}}_{23}$ (indicating coincidences between detectors $D_{H1}$ and $D_{V2}$ or between detectors $D_{V1}$ and $D_{H2}$) and $\ket{\Phi^{-}}_{23}$ (indicating coincidences between detectors $D_{H1}$ and $D_{H2}$ or between detectors $D_{V1}$ and $D_{V2}$). Thus, if there are coincidences between detectors $D_{1}$ ($D_{H1}$ or $D_{V1}$) and $D_{2}$ ($D_{H2}$ or $D_{V2}$), then 1-bit trigger information will be sent to do the related Pauli operations on the outputs, and we will get the desired heralded CNOT gate. The total success probability is 1/8 \cite{supp}, which can be improved to a optimal value of 1/4 by harnessing complete Bell-state analyser with more photons \cite{grice2011arbitrarily, ewert20143} and hybrid degree of freedoms \cite{fiorentino2004deterministic}.

Note that when there are multiple photons or no photon in path 2 or 3, unwanted coincidences might happen between detectors $D_{1}$ ($D_{H1}$ or $D_{V1}$) and $D_{2}$ ($D_{H2}$ or $D_{V2}$), leading to uncorrected CNOT operations. However, these cases can be excluded with photon-number-resolving detectors (PNRDs) \cite{fitch2003photon, nehra2020photon} and photon bunching effect \cite{eisenberg2005multiphoton} provided by PBS3 in $\ket{R}/\ket{L}$ basis (please refer to \cite{supp} for details). For a high-performance heralded CNOT gate, a truly on-demand SPS together with PNRDs will be the key resources. Here, we demonstrate it by using the best all-around QD-based SPS and pseudo-PRNDs constructed by multiple superconducting nanowire single-photon detectors (SNSPDs).

\begin{figure*}[!t]
	\includegraphics[width=0.88\textwidth]{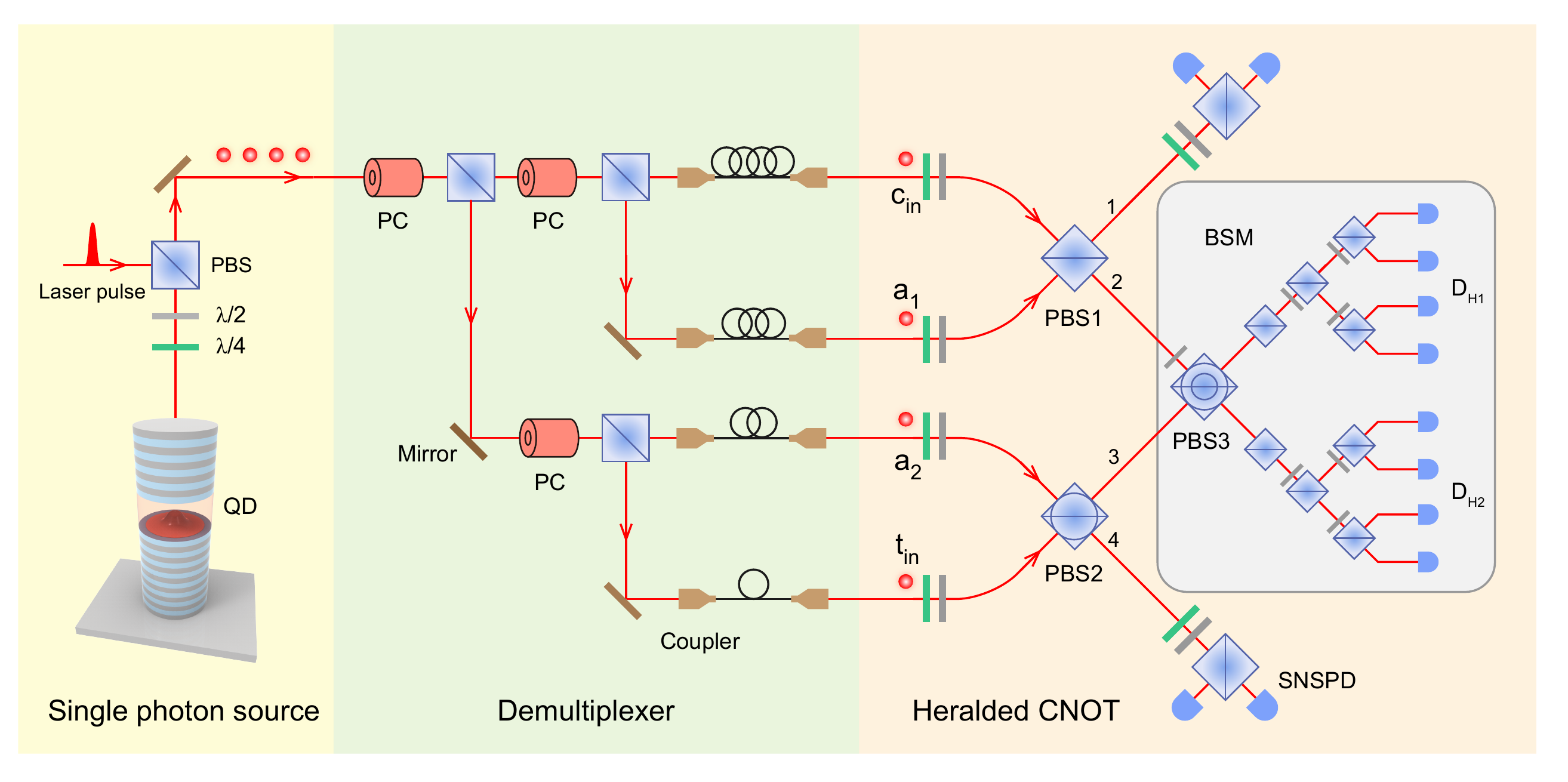}
	\caption{The experimental setup. A single InAs/GaAs quantum dot  (QD), resonantly coupled to a microcavity, is used to create pulsed resonance fluorescence single photons. For demultiplexing, three pairs of Pockels cells (PCs, extinction ratio $>$100:1) and PBSs  (extinction ratio $>$2000:1) are used to actively translate a stream of single photons into four spatial modes. Four single-mode fibers with different lengths (each fiber output port is mounted on a translation stage, without drawn), are used to precisely compensate time delays. Then we adjust the wave-plates to prepare general input states, and polarized states $\ket{+}$ and $\ket{H}$ for ancillary photons as described in Fig.\ref{fig:Concept}. These single photons are fed into the heralded CNOT gate. We perform a jointly projective measurement of the Bell state $\ket{\Phi^{+}}$ (a half wave-plate between PBS2 and PBS3 was setted at zero degree for transforming $\ket{\Phi^{+}}$ to  $\ket{\Phi^{-}}$), yielding the desired output state of the CNOT operation. The wave-plates and a PBS are used for making the polarization basis projection. The single photons are then detected by pseudo-PNRDs constructed by SNSPDs. A successful heralded CNOT operation will be achieved if there is a coincidence between PNRDs $D_{H1}$ and $D_{H2}$ (or $D_{V1}$ and $D_{V2}$, not show). All four-fold coincidences are recorded by a multi-channel coincidence count unit for estimating the gate fidelity \cite{supp}.}
	\label{fig:Setup}
\end{figure*}

As shown in Fig.\ref{fig:Setup}, we use the state-of-the-art self-assembled InAs/GaAs QDs embedded inside a micro-pillar cavity \cite{unsleber2016highly} to create single photons of near-perfect purity, indistinguishability and high brightness. To reach the best QD-cavity coupling with optical resonance $\sim$893nm, the whole sample wafer was mounted in an ultra-stable liquid-helium-free bath cryostat and cooled down to 4K. Under pulse resonant excitation with a laser repetition rate $\sim$76MHz, $\sim$6.4 MHz polarized resonance fluorescence single photons are directly registered by a single-mode fiber coupled SNSPD with a detector efficiency $\sim$80\% (without any filters). The measured second-order correlation function at zero-time delay is $0.03(1)$, yielding single-photon purity of $\sim$97\%. The photon indistinguishability is measured by a Hong-Ou-Mandel interferometer, yielding a visibility of 0.91(1) between two photons separated by $\sim6.5\mu s$ \cite{li2020multiphoton}.  

The produced single-photon stream is then deterministically demultiplexed into four spatial modes using the demultiplexer constructed by three pairs of Pockels cells (PCs) and PBSs which are customized for the SPS \cite{supp,wang2017high}. These PCs will actively control the photon polarization when loaded with high-voltage electrical pulses, synchronized to the laser pulses and operated at a repetition rate $\sim$0.76MHz. That means in each operation cycle, every 100 single photons will be divided equally into four paths \cite{supp}. Thanks to the high transmission efficiency ($>$99\%) and high single-mode fiber coupling efficiency ($\sim$85\%), we can reach the average optical switches efficiency $\sim$83\% \cite{supp, wang2017high}. By using single-mode fibers of different lengths and mounting each fiber output end on a translation stage, we can precisely compensate time delays of the four single photons that are fed into the CNOT gate.

Our heralded CNOT operation depends on the ancilla measurement outcome and their related Pauli operators. For simplicity, we perform a jointly projective measurement of Bell state $\ket{\Phi^{+}}$ on ancillary photons in path 2 and 3, indicating that the output state is exactly the outcome of a CNOT gate. As shown in Fig.\ref{fig:Setup}, $\ket{\Phi^{+}}$ corresponds to the coincidence between detectors $D_{H1}$ and $D_{H2}$ or between detectors $D_{V1}$ and $D_{V2}$ (not show). We note that unwanted coincidences between heralded detectors can be excluded by pseudo-PNRDs and photon bunching effect \cite{eisenberg2005multiphoton,supp}.

\begin{figure*}[!t]
	\includegraphics[width=0.97\textwidth]{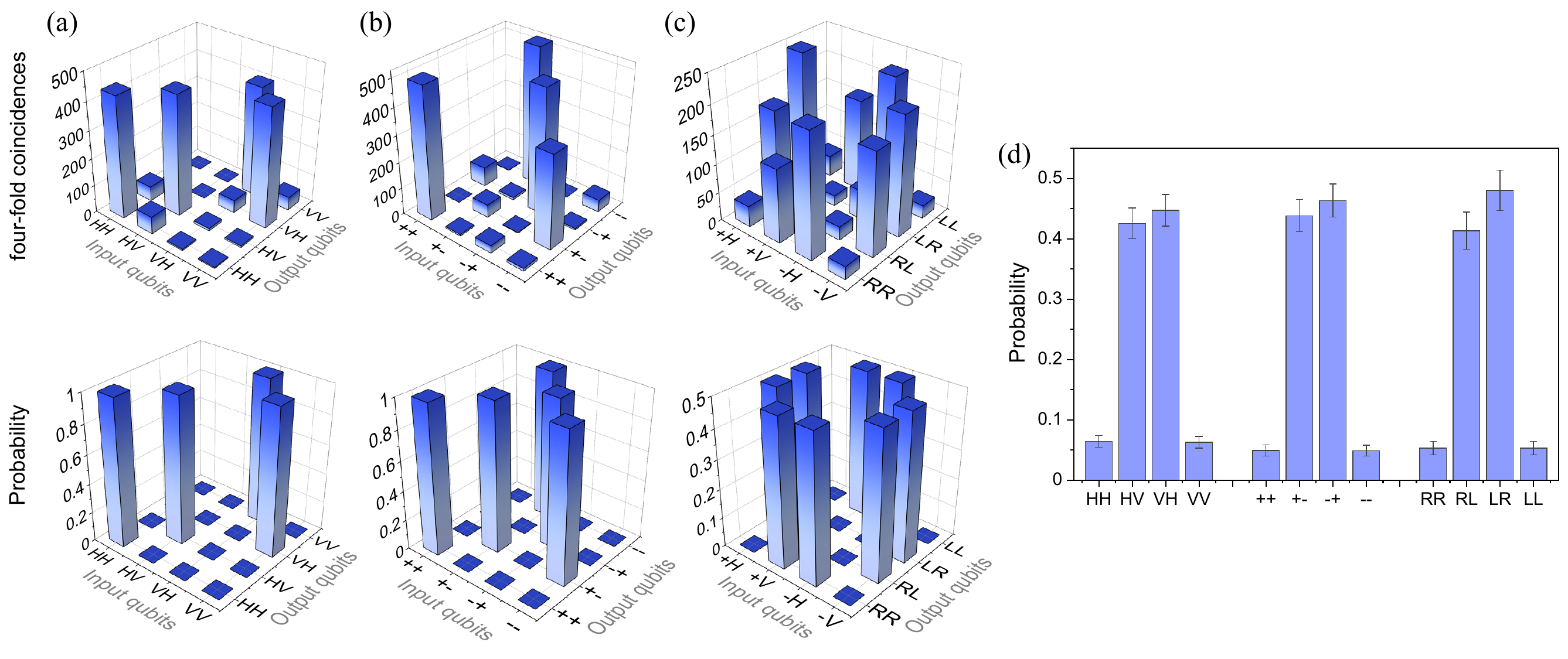}
	\caption{Experimentally achieved heralded CNOT gate and event-ready Bell states (collected in 6 minutes). Up: four-fold coincidences for all possible combinations of inputs and outputs; Down: ideal gates with 100\% fidelity. \textbf{(a)}: In $\ket{H}/\ket{V}$ basis. \textbf{(b)}: In $\ket{+}/\ket{-}$ basis. \textbf{(c)}: We measure the output qubits in $\ket{R}/\ket{L}$ basis when the control and target qubits are in $\ket{+}/\ket{-}$ and $\ket{H}/\ket{V}$ basis. \textbf{(d)}:  Bell state $\ket{\Psi^{-}}$ produced by the heralded CNOT gate for input state $\ket{-}_{c}\ket{V}_{t}$. The coincidence counts are measured in mutually unbiased bases (here we show the probabilities).} 
	\label{fig:AllResults}
\end{figure*}

To experimentally evaluate the CNOT operation, we exploit an efficient approach proposed by Hofmann \cite{hofmann2005complementary}. We prepare the input qubits in the computation basis ($\ket{H}/\ket{V}$) as well as the superposition states ($\ket{+}/\ket{-}$). Then we measure the coincidence counts of all possible combinations in each basis recorded by a home-made multi-channel coincidence count unit, as summarized in Fig.\ref{fig:AllResults}(a) and (b). We can see that the gate works well in both bases and the achieved experimental fidelities (defined as the probability of obtaining the correct output averaged over all four possible inputs) are estimated to be $F_{1}=(87.8\pm2.1)\%$ in $\ket{H}/\ket{V}$ basis and $F_{2}=(88.6\pm2.1)\%$ in $\ket{+}/\ket{-}$ basis \cite{supp}. With the two complementary fidelities $F_{1}$ and $F_{2}$, we can bound the quantum process fidelity $F_{proc}$ according to $\left(F_{1}+F_{2}-1\right)\leq F_{proc} \leq min\left(F_{1},F_{2}\right)$, yielding ($76.4\pm2.9)\%\leq F_{proc} \leq(87.8\pm2.1$)\%. The process fidelity $F_{proc}$ is directly related to the entangling capability, which means it can produce entangled states from separable states. Our result is clearly over the threshold of 0.5 that is sufficient to confirm the gate's entangling ability \cite{hofmann2005complementary}. Moreover, if the average fidelity of three distinct classical operations exceeds 2/3, one can say that local operations and classical communications cannot reproduce the gate \cite{hofmann2005quantum}. We demonstrate it by preparing the control input in $\ket{+}/\ket{-}$ basis and target input in $\ket{H}/\ket{V}$ basis, and performing the measurement on output qubits in $\ket{R}/\ket{L}$ basis. The experimental result is shown in Fig.\ref{fig:AllResults}(c), which gives fidelity $F_{3}=(87.0\pm2.2)\%$. The average gate fidelity of $F_{1}$, $F_{2}$ and $F_{3}$ is $(87.8\pm1.2)\%$, obviously exceeding the boundary of 2/3. 

\begin{table*}[!t]
	\centering
	\caption{A comparison of linear optical CNOT gates (for more, see \cite{supp}). SPSs: single-photon sources; PNRDs: photon-number resolving detectors; $P_{s}$: theoretical success probability; $\eta_{h}$: heralding efficiency; /: not exit; --: unreported.}
	\begin{tabular}{|c|c|c|P{2cm}|P{1.5cm}|P{1.5cm}|c|}
		\hline 
		experiments &on-demand SPSs& (pseudo) PNRDs & Heralded &$P_{s}$&$\eta_{h}$&Operation/min\\\hline 
		O'Brien \textit{et al.}\cite{o2003demonstration}&no&no&no&1/9&/&$<1$\\ \hline 
		Okamoto \textit{et al.}\cite{okamoto2011realization}&no&no&no&1/16&/&$<2$\\ \hline
		Gasparoni \textit{et al.}\cite{gasparoni2004realization}&no&no&no$^{*}$&1/4&$<10^{-4}$&$<5$\\ \hline  	  
		Bao \textit{et al.}\cite{bao2007optical}&no&no&no$^{*}$&1/8&$<10^{-5}$&$<1$\\ \hline 
		He \textit{et al.}\cite{he2013demand}&yes&no&no&1/9&/& --\\ \hline 
		Gazzano \textit{et al.}\cite{gazzano2013entangling}&yes&no&no&1/9&/&-- \\\hline  
		This Work &yes&yes&yes&1/8&$\sim$0.008&$\sim$85 \\ \hline
		\multicolumn{7}{l}{$^{*}$\footnotesize{measurement of output states for postselection due to the multiple pair emission of SPDC.}} \\  
	\end{tabular}
	\label{tab:comparisons}
\end{table*}  
As an application of the heralded CNOT gate, we produce event-ready entangled states by preparing a separable state $\ket{-}_{c}\ket{V}_{t}$ at the inputs. Corresponding to the CNOT operation, we expect outputting a maximally entangled Bell state $\ket{\Psi^{-}}=1/\sqrt{2}\left(\ket{HV}-\ket{VH}\right)$. To verify that it was implemented successfully, we measured the correlation between the polarizations of control and target photons in different basis, as shown in Fig.\ref{fig:AllResults}(d). The average fidelity of the produced state is $F_{\Psi^{-}}=(83.4\pm 2.4)\%$, which clearly surpassed the classical threshold of 0.5 \cite{guhne2009entanglement, wang2016experimental} and ensures entanglement. Moreover, we can achieve $\sim$85 CNOT operations per minute (defined as all output four-fold coincidence counts over the collected time for the input state), which is increased by at least an order of magnitude compared to previous experiments in Table.\ref{tab:comparisons}.

Our demonstrated CNOT gate is heralded and has a high heralding efficiency that could be up to one in principle. What we mean heralding efficiency is, for a given input state, the probability of achieving a desired CNOT gate when there are coincidences between heralded detectors. For simplicity, we define the experimental heralding efficiency $\eta_{h}$ as the ratio between probabilities of four-fold and two-fold coincidence counts\cite{supp}. One can achieve $\eta_{h}\approx0.008$ in the experiment due to the imperfect single photon efficiency $\eta_{s}$ ($\eta_{s}=\eta_{f}\eta_{w}\eta_{l}$, $\eta_{f}$, $\eta_{w}$ and $\eta_{l}$ are the efficiencies for photon brightness at the fiber output of confocal system, switches and optical line) when detector efficiency $\eta_{d}=0.8$ \cite{supp}. However, it is still increased by at least an order of magnitude compared to previous experiments. Also, by coupling QDs to an asymmetric micro-cavity \cite{wang2019towards}, the output brightness $\eta_{f}$ of our QD SPS will hopefully reach to near-unity. Moreover, the switches efficiency can gradually increase close to one in principle \cite{supp}, which also helps improving $\eta_{s}$. With ideal $\eta_{s}$, $\eta_{h}$ could reach to one using perfect PNRDs \cite{supp}. Additionally, the indistinguishability and purity of QD-based SPSs can be both improved to near-unity \cite{senellart2017high, he2013demand}, indicating that the fidelity of CNOT gates can also reach to near-unity and the count rates can be further greatly extended.

In summary, by using high-quality single photons produced from QD-micropillar based on-demand SPSs and pseudo-PNRDs constructed by SNSPDs, we have for the first time implemented an optical heralded CNOT gate of high fidelity, high rates and heralding efficiency increased by at least an order of magnitude. Our results are promising for various QIP tasks such as complete Bell state analysis in quantum teleportation \cite{bennett1993teleporting, bouwmeester1997experimental} and heralded creation of multiphoton entanglement especially cluster states \cite{raussendorf2001one, lindner2009proposal, schwartz2016deterministic}, which is important for large-scale quantum computation. 

Interestingly, the single photons for our CNOT gates can be generated by separate sources that can be far away. Then one can realize remote entanglement generation and quantum gates, which are useful for distributed quantum computing and will find new applications in the future quantum internet. Furthermore, our system can be incorporated in a realistic fiber systems \cite{clark2009all, chen2008demonstration} using QD SPSs at telecommunication band \cite{portalupi2019inas,weber2019two}, thus one can further explore long-distance quantum communication and fibre-optic quantum network. Finally, we suggest that recent developments of integrated optics \cite{wang2019integrated} could be particularly useful to fully realize the demonstrated experiment for miniaturized and scalable photonic QIP.

\begin{acknowledgements}
This work was supported by the National Natural Science Foundation of China, the Chinese Academy of Sciences, and the National Basic Research Program of China (973 Program).

Jin-Peng Li, Xuemei Gu and Jian Qin contributed equally to this work.
\end{acknowledgements}

\bibliographystyle{unsrt}
\bibliography{refs}

\newpage 

\onecolumngrid
\section{Supplementary Information for Heralded non-destructive quantum entangling gate with single-photon sources}

\textbf{This Supplementary file includes:}
\vspace{0.1cm}

\normalsize
\hyperlink{part 01}{1. Detailed work principle of the heralded CNOT gate}
\vspace{0.05cm}

\hspace{0.2cm}\hyperlink{part 01-1}{(1). the desired case for heralded CNOT gates: only one photon in each path 1, 2, 3 and 4}

\hspace{0.2cm}\hyperlink{part 01-2}{(2). undesired cases which need to be excluded}

\vspace{0.05cm}
\hyperlink{part 02}{2. Fidelity calculation for the heralded CNOT gate and event-ready Bell state}

\hspace{0.2cm}\hyperlink{part 02-1}{(1). Fidelity calculation for the heralded CNOT gate}

\hspace{0.2cm}\hyperlink{part 02-2}{(2). Fidelity calculation for the event-ready Bell state}

\vspace{0.05cm}
\hyperlink{part 03}{3. Detailed analysis of the heralding efficiency}

\vspace{0.05cm}
\hyperlink{part 04}{4. The details of the demultiplexer}

\vspace{0.05cm}
\hyperlink{part 05}{5. A comparison of experimental linear optical CNOT gates}

\hyperlink{part 06}{References for Supplementary Information reference citations}

\hypertarget{part 01}{\subsection{1. Detailed work principle of the heralded CNOT gate}}

A CNOT gate flips the second (target) bit if and only if the first one (control) has the logical value 1 and the control bit remains unaffected. In our experiment, the logical value of each of the qubits is represented by the polarization state of a single photon, where a horizontal polarization state $\ket{H}$ stands for a logical value $0$ and vertical polarization state $\ket{V}$ represents the logical value $1$. Thereby, the logic table of the CNOT operation is described as following:
\begin{center}
	$\ket{H}_{c_{in}}\ket{H}_{t_{in}}\rightarrow\ket{H}_{c_{out}}\ket{H}_{t_{out}}$,\hspace{0.5cm}$\ket{H}_{c_{in}}\ket{V}_{t_{in}}\rightarrow\ket{H}_{c_{out}}\ket{V}_{t_{out}}$,\\
	$\ket{V}_{c_{in}}\ket{H}_{t_{in}}\rightarrow\ket{V}_{c_{out}}\ket{V}_{t_{out}}$,\hspace{0.5cm}$\ket{V}_{c_{in}}\ket{V}_{t_{in}}\rightarrow\ket{V}_{c_{out}}\ket{H}_{t_{out}}$
\end{center}

To implement a heralded nondestructive CNOT gate, we exploit the scheme that requires only four independent single photons \cite{bao2007optical}. As described in Fig.1 in the main text, the scheme performs a heralded CNOT operation on the input photons in spatial modes $c_{in}$ and $t_{in}$. The output qubits are contained in spatial modes $c_{out}$ and $t_{out}$ and the ancillary photons in the spatial modes $a_{1}$ and $a_{2}$ are in the states of $\ket{\psi}_{a_{1}}=1/\sqrt{2}\left(\ket{H}+\ket{V}\right)$ and $\ket{\psi}_{a_{2}}=\ket{H}$. For the gate to work properly, we assume the most general input states $\ket{\psi}_{c_{in}}=\alpha\ket{H}+\beta\ket{V}$ and $\ket{\psi}_{t_{in}}=\gamma\ket{H}+\delta\ket{V}$, where the complex coefficients $\alpha$, $\beta$, $\gamma$ and $\delta$ satisfy $|\alpha|^2+|\beta|^2=1$ and $|\gamma|^2+|\delta|^2=1$. Therefore, the total state of the input four photons can be expressed as: 
\begin{align}
\ket{\Psi_{in}}=\nonumber&\ket{\psi}_{c_{in}}\ket{\psi}_{a_{1}}\ket{\psi}_{a_{2}}\ket{\psi}_{t_{in}} \\\nonumber
=&\Big(\alpha\ket{H}+\beta\ket{V}\Big)_{c_{in}}\otimes\frac{1}{\sqrt{2}}\Big(\ket{H}+\ket{V}\Big)_{a_{1}} \otimes\Big(\ket{H}\Big)_{a_{2}}\otimes\Big(\gamma\ket{H}+\delta\ket{V}\Big)_{t_{in}} \\
=&\Big(\alpha\ket{H}+\beta\ket{V}\Big)_{c_{in}}\otimes\frac{1}{\sqrt{2}}\Big(\ket{H}+\ket{V}\Big)_{a_{1}}\otimes\frac{1}{\sqrt{2}}\Big(\ket{+}+\ket{-}\Big)_{a_{2}}\otimes\frac{1}{\sqrt{2}}\Big(\gamma(\ket{+}+\ket{-})+\delta(\ket{+}-\ket{-})\Big)_{t_{in}}
\label{eq:CNOTinstate}
\end{align}
where the subscript stands for the photon's path, and states $\ket{+}$ equals to $1/\sqrt{2}\left(\ket{H}+\ket{V}\right)$ and $\ket{-}$ is $1/\sqrt{2}\left(\ket{H}-\ket{V}\right)$.

Let the ancillary photon in path $a_{1}$ and the control photon in path $c_{in}$ superimpose at the polarizing beam splitter PBS1 in $\ket{H}/\ket{V}$ basis, which transmits horizontal polarization state $\ket{H}$ and reflects vertical polarization state $\ket{V}$. Consequently, the input state $\ket{\Psi}_{in}$ at the output ports of the PBS1 becomes:
\begin{align}
\ket{\Psi}_{in}\nonumber\xrightarrow{PBS1}\ket{\Psi}_{PBS1}   \nonumber
&=\Big(\alpha\ket{H}_{2}+\beta\ket{V}_{1}\Big)\frac{1}{\sqrt{2}}\Big(\ket{H}_{1}+\ket{V}_{2}\Big) \\ \nonumber
&\otimes\frac{1}{\sqrt{2}}\Big(\ket{+}+\ket{-}\Big)_{a_{2}}\frac{1}{\sqrt{2}}\Big(\gamma(\ket{+}+\ket{-})+\delta(\ket{+}-\ket{-})\Big)_{t_{in}}
\end{align}
where subscript 1 and 2 represent the photon's path after PBS1.

Similarly, after passing through PBS2 in $\ket{+}/\ket{-}$ basis that transmits $\ket{+}$ state and reflects $\ket{-}$ one (PBS2 is constructed with a PBS in $\ket{H}/\ket{V}$ basis and four half-wave plates (HWP)), the state of the two photons in paths $t_{in}$ and $a_{2}$ converts into
\begin{align}
\ket{\Psi}_{PBS1}\xrightarrow{PBS2}\ket{\Psi}_{PBS1,2}    \nonumber
&=\Big(\alpha\ket{H}_{2}+\beta\ket{V}_{1}\Big)\frac{1}{\sqrt{2}}\Big(\ket{H}_{1}+\ket{V}_{2}\Big) \\ \nonumber
&\otimes\frac{1}{\sqrt{2}}\Big(\ket{+}_{4}+\ket{-}_{3}\Big)\frac{1}{\sqrt{2}}\Big(\gamma(\ket{+}_{3}+\ket{-}_{4})+\delta(\ket{+}_{3}-\ket{-}_{4})\Big)
\end{align}
where subscript 3 and 4 represent the photon's path after PBS2. Thereby, the state after PBS1 and PBS2 can be expanded and rewritten as following:
\begin{align}
\ket{\Psi}_{PBS1,2}=\nonumber&\frac{1}{\sqrt{2}}\Big(\alpha\ket{H}_{1}\ket{H}_{2}+\alpha\ket{H}_{2}\ket{V}_{2}+\beta\ket{H}_{1}\ket{V}_{1}+\beta\ket{V}_{1}\ket{V}_{2}\Big)\\ \nonumber
&\otimes\frac{1}{2}\Big[\gamma(\ket{+}_{3}\ket{+}_{4}+\ket{+}_{4}\ket{-}_{4}+\ket{-}_{3}\ket{+}_{3}+\ket{-}_{3}\ket{-}_{4}) \\
&\quad+\delta(\ket{+}_{3}\ket{+}_{4}-\ket{+}_{4}\ket{-}_{4}+\ket{-}_{3}\ket{+}_{3}-\ket{-}_{3}\ket{-}_{4})\Big]
\label{eq:outputState}
\end{align}

The successful CNOT operation is heralded by the measurement of additional ancillary photons in path 2 and 3, and the output state of photons in path $c_{out}$ and $t_{out}$ describe the result of CNOT gates as shown in Fig.1 of the main text; This corresponds to the case that only one photon in each path after PBS1 and PBS2. However, from the Eq.(\ref{eq:outputState}), we find that there can be more than one photon in few path such as the term $\ket{H}_{2}\ket{V}_{2}\ket{+}_{3}\ket{-}_{3}$ or no photon in certain paths. We now discuss all possible cases in the next text.

\hypertarget{part 01-1}{\subsection{(1). the desired case for heralded CNOT gates: only one photon in each path 1, 2, 3 and 4}}

Let us consider the case that there is only one photon in each output port after PBS1 and PBS2. As a result, the efficient four-photon state is given by:
\begin{align}
\ket{\Psi} 
=&\Big(\alpha\ket{H}_{1}\ket{H}_{2}+\beta\ket{V}_{1}\ket{V}_{2}\Big)\otimes\frac{1}{\sqrt{2}}\Big[\gamma(\ket{+}_{3}\ket{+}_{4}+\ket{-}_{3}\ket{-}_{4})+(\delta\ket{+}_{3}\ket{+}_{4}-\ket{-}_{3}\ket{-}_{4})\Big] \label{eq:eachmodeonephoton} \\
=&\Big(\alpha\ket{H}_{1}\ket{H}_{2}+\beta\ket{V}_{1}\ket{V}_{2}\Big)\otimes\frac{\gamma}{\sqrt{2}}\Big(\ket{H}_{3}\ket{H}_{4}+\ket{V}_{3}\ket{V}_{4}\Big)
\label{eq:swapping01} \\
&+\Big(\alpha\ket{H}_{1}\ket{H}_{2}+\beta\ket{V}_{1}\ket{V}_{2}\Big)\otimes\frac{\delta}{\sqrt{2}}\Big(\ket{H}_{3}\ket{V}_{4}+\ket{V}_{3}\ket{H}_{4}\Big)
\label{eq:swapping02}
\end{align}
with a probability of 1/4 from Eq.(\ref{eq:outputState}). From Eq.(\ref{eq:swapping01}) (or Eq.(\ref{eq:swapping02})), we can see that, when one does a quantum swapping between these two entangled states (which means projecting photons in path 2 and 3 into one of four Bell states), one will get the quantum state of photons in path 1 and 4, which is equivalent to the result of a CNOT operation between input states $\ket{\psi}_{c_{in}}=\alpha\ket{H}+\beta\ket{V}$ and $\gamma\ket{H}_{t_{in}}$ (or $\delta\ket{V}_{t_{in}}$) with relative single-qubit errors. We will explain them later in this section. The concept of entanglement swapping partially determined the states of ancillary photons that are needed, since we need to construct entangled states as described in Eq.(\ref{eq:swapping01}) and Eq.(\ref{eq:swapping02}) and then do the quantum swapping. Notice that when two ancillary photons are in states $\ket{\psi}_{a_{1}}=1/\sqrt{2}\left(\ket{H}\pm\ket{V}\right)$ and $\ket{\psi}_{a_{2}}=\ket{H}$ (or $ \ket{V}$), desired heralded CNOT gates can also be obtained with different single-qubit error correction if we project states into the same Bell state. For simplicity, we use the ancillary photons in the states of $\ket{\psi}_{a_{1}}=1/\sqrt{2}\left(\ket{H}+\ket{V}\right)$ and $\ket{\psi}_{a_{2}}=\ket{H}$ for the detailed derivation.

Now let's expand Eq.(\ref{eq:eachmodeonephoton}) in the Bell basis as follows:
\begin{align}
\ket{\Psi}&=\nonumber\frac{1}{\sqrt{2}}\Big[\alpha(\ket{\Phi^{+}}_{12}+\ket{\Phi^{-}}_{12})+\beta(\ket{\Phi^{+}}_{12}-\ket{\Phi^{-}}_{12})\Big]\otimes\Big(\gamma\ket{\Phi^{+}}_{34}+\delta\ket{\Psi^{+}}_{34}\Big)\\\nonumber
&=\frac{1}{\sqrt{2}}\Big[(\alpha+\beta)\ket{\Phi^{+}}_{12}+(\alpha-\beta)\ket{\Phi^{-}}_{12}\Big]\otimes\Big(\gamma\ket{\Phi^{+}}_{34}+\delta\ket{\Psi^{+}}_{34}\Big)\\\nonumber
&= \frac{1}{{\sqrt{2}}}\Big(\gamma[(\alpha+\beta)\ket{\Phi^{+}}_{12}\ket{\Phi^{+}}_{34})+(\alpha-\beta)\ket{\Phi^{-}}_{12}\ket{\Phi^{+}}_{34}]+\delta[(\alpha+\beta)\ket{\Phi^{+}}_{12}\ket{\Psi^{+}}_{34}+(\alpha-\beta)\ket{\Phi^{-}}_{12}\ket{\Psi^{+}}_{34}]\Big) \\
&=\frac{1}{\sqrt{2}}\Big(\gamma[(\alpha+\beta)+(\alpha-\beta)\sigma_{z1}]+\delta[(\alpha+\beta)\sigma_{x4}+(\alpha-\beta)\sigma_{z1}\sigma_{x4}]\Big)\ket{\Phi^{+}}_{12}\ket{\Phi^{+}}_{34}
\label{eq:outputBellState}
\end{align}
where $\ket{\Phi^{\pm}}=1/\sqrt{2}(\ket{H}\ket{H}\pm\ket{V}\ket{V})$ and $\ket{\Psi^{\pm}}=1/\sqrt{2}(\ket{H}\ket{V}\pm\ket{V}\ket{H})$ are standard Bell states in $\ket{H}/\ket{V}$ basis. $\sigma_{z1}$ and $\sigma_{x4}$ are Pauli Z and X operators on photons in path 1 and 4.

We further expand the state in Eq.(\ref{eq:outputBellState}) with the Bell states of photon 2 and 3, yielding:
\begin{align}
\ket{\Phi^{+}}_{12}\ket{\Phi^{+}}_{34}\nonumber&=\frac{1}{2}\Big(\ket{\Phi^{+}}_{14}\ket{\Phi^{+}}_{23}+\ket{\Phi^{-}}_{14}\ket{\Phi^{-}}_{23}+\ket{\Psi^{+}}_{14}\ket{\Psi^{+}}_{23}+\ket{\Psi^{-}}_{14}\ket{\Psi^{-}}_{23}\Big)\\
&=\frac{1}{2}\Big(\ket{\Phi^{+}}_{14}\ket{\Phi^{+}}_{23}+\sigma_{z1}\ket{\Phi^{+}}_{14}\ket{\Phi^{-}}_{23}+\sigma_{x4}\ket{\Phi^{+}}_{14}\ket{\Psi^{+}}_{23}+\sigma_{z1}\sigma_{x4}\ket{\Phi^{-}}_{14}\ket{\Psi^{-}}_{23}\Big)
\label{eq:outputBellState2}
\end{align}

We then substitute $\ket{\Phi^{+}}_{12}\ket{\Phi^{+}}_{34}$ in Eq.(\ref{eq:outputBellState2}) into Eq.(\ref{eq:outputBellState}), which leads to:
\begin{align}
\ket{\Psi}=&\frac{1}{2}\Big(\ket{\psi}_{14}^{out}\ket{\Phi^{+}}_{23}+\sigma_{z1}\ket{\psi}_{14}^{out}\ket{\Phi^{-}}_{23}+\sigma_{x4}\ket{\psi}_{14}^{out}\ket{\Psi^{+}}_{23}+\sigma_{z1}\sigma_{x4}\ket{\psi}_{14}^{out}\ket{\Psi^{-}}_{23}\Big)
\label{eq:expanededCNOT}
\end{align}
where $\ket{\psi}_{14}^{out}=1/\sqrt{2}\gamma\Big[(\alpha+\beta)\ket{\Phi^{+}}_{14}+(\alpha-\beta)\ket{\Phi^{-}}_{14}\Big]+1/\sqrt{2}\delta\Big[(\alpha+\beta)\ket{\Psi^{+}}_{14}+(\alpha-\beta)\ket{\Psi^{-}}_{14}\Big]$, and $\ket{\psi}_{14}^{out}$ represents the output state of photons in path 1 and 4, which is equivalent to the result of a CNOT gate operation on input state $\ket{\psi}_{c_{in}}\ket{\psi}_{t_{in}}$. We now demonstrate the statement. 

In our scheme, the essentially standard CNOT operation between photons in path $c_{in}$ and $t_{in}$ are defined as $U_{14}$, which can be written as following:
\begin{align}
U_{14}\ket{\psi}_{c_{in}}\ket{\psi}_{t_{in}}\nonumber
&=U_{14}\Big(\alpha\ket{H}+\beta\ket{V}\Big)_{c_{in}}\otimes\Big(\gamma\ket{H}+\delta\ket{V}\Big)_{t_{in}} \\\nonumber
&=U_{14}\Big(\alpha\ket{H}+\beta\ket{V}\Big)_{c_{in}}\otimes\gamma\ket{H}_{t_{in}}+U_{14}\Big(\alpha\ket{H}+\beta\ket{V}\Big)_{c_{in}}\otimes\delta\ket{V}_{t_{in}} \\\nonumber
&=\Big[\gamma(\alpha\ket{H}\ket{H}+\beta\ket{V}\ket{V})+\delta(\alpha\ket{H}\ket{V}+\beta\ket{V}\ket{H})\Big]_{c_{in},t_{in}} \\\nonumber
&=\frac{1}{\sqrt{2}}\Big(\gamma\Big[\alpha(\ket{\Phi^{+}}+\ket{\Phi^{-}})+\beta(\ket{\Phi^{+}}-\ket{\Phi^{-}})\Big]+\delta\Big[\alpha(\ket{\Psi^{+}}+\ket{\Psi^{-}})+\beta(\ket{\Psi^{+}}-\ket{\Psi^{-}})\Big]\Big)_{c_{in},t_{in}}\\
&=\frac{1}{\sqrt{2}}\Big(\gamma\Big[(\alpha+\beta)\ket{\Phi^{+}}_{14}+(\alpha-\beta)\ket{\Phi^{-}}_{14}\Big]+\delta\Big[(\alpha+\beta)\ket{\Psi^{+}}_{14}+(\alpha-\beta)\ket{\Psi^{-}}_{14}\Big]\Big)_{c_{in},t_{in}}
\label{eq:CNOTgate}
\end{align} 

Compare the Eq.(\ref{eq:CNOTgate}) and $\ket{\psi}_{14}^{out}$, we can say that they are the same. That means the results of CNOT operation between photons in path $c_{in}$ and $t_{in}$ is equivalent to the output of photons from paths 1 ($c_{out}$) and 4 ($t_{out}$). Thus, we have $U_{14}\ket{\psi}_{c_{in}}\ket{\psi}_{t_{in}}=\ket{\psi}_{14}^{out}$ and one can substitute Eq.(\ref{eq:CNOTgate}) to Eq.(\ref{eq:expanededCNOT}), which finally leads to the equation described in the main text: 
\begin{align}
\ket{\Psi}=\nonumber& \frac{1}{2}\Big[I_{1}I_{4}U_{14}\ket{\psi}_{1}^{c_{in}}\ket{\psi}_4^{t_{in}}\otimes\ket{\Phi^{+}}_{23}+I_{1}\sigma_{x4}U_{14}\ket{\psi}_{1}^{c_{in}}\ket{\psi}_4^{t_{in}}\otimes\ket{\Psi^{+}}_{23}\\
&+ \sigma_{z1}I_{4}U_{14}\ket{\psi}_{1}^{c_{in}}\ket{\psi}_4^{t_{in}}\otimes\ket{\Phi^{-}}_{23}+\sigma_{z1}\sigma_{x4}U_{14}\ket{\psi}_{1}^{c_{in}}\ket{\psi}_4^{t_{in}}\otimes\ket{\Psi^{-}}_{23}\Big]
\label{eq:CNOToutputstate}
\end{align}
where $I_{1}$($I_{4}$) is the identity operation on the photon in path 1 (4).

From the Eq.(\ref{eq:CNOToutputstate}), we can see that if the jointly measurement result of photons in path 2 and 3 is the Bell state $\ket{\Phi^{+}}_{23}$, then the state of photons in path 1 and 4 is exactly the output state of the CNOT operation; if the jointly measurement result is another Bell state (such as $\ket{\Phi^{-}}_{23}$, $\ket{\Psi^{+}}_{23}$ or $\ket{\Psi^{-}}_{23}$), then corresponding single-qubit operations on the state of photons in path 1 and 4 are required to get the result of the desired CNOT gate. However, for standard linear optical Bell-state analyser without ancillary photons, only two of the four Bell states can be distinguished \cite{calsamiglia2001maximum}. Therefore, the heralded non-destructive CNOT gate is successfully implemented with a total success probability$$P_{success}=\frac{1}{4}\times\frac{1}{2}=\frac{1}{8},$$ where number $1/4$ ($1/2$) represents the probability of getting the case that each path has only one photon after photons pass through PBS1 and PBS2 (Bell states measurement). The $P_{success}$ can be further improved by harnessing complete Bell-state analyser assisted with more ancillary photons \cite{grice2011arbitrarily,ewert20143} and hybrid degree of freedoms \cite{fiorentino2004deterministic}.

\hypertarget{part 01-2}{\subsection{(2). undesired cases which can be excluded}}

Until now, we have discussed the expected case that there is only one photon in each output port of PBS1 and PBS2. However, as described in Eq.(\ref{eq:outputState}), other cases that a path contains two or zero photons also exist. Considering PBS1 and PBS2 jointly, there will be nine cases in three groups as follows \cite{bao2007optical}:

\hspace{1cm}Group 1: \hspace{0.2cm} 1:1:1:1,

\hspace{1cm}Group 2: \hspace{0.2cm} 2:0:2:0, \hspace{0.1cm} 0:2:0:2,

\hspace{1cm}Group 3: \hspace{0.2cm} 1:1:2:0, \hspace{0.2cm}1:1:0:2, \hspace{0.2cm} 2:0:1:1, \hspace{0.2cm} 0:2:1:1, \hspace{0.2cm} 2:0:0:2, \hspace{0.2cm} 0:2:2:0\\
where $n_{1}:n_{2}:n_{3}:n_{4}$ corresponds to the photon numbers in each path (1, 2, 3 and 4) as described in Fig.1 of the main text. Group 1 is exactly what we want, where the CNOT operation is successfully implemented, just as what we have discussed early. 

For the cases in Group 2, there are totally two photons in path 2 and 3. However, the two photons are bunching into the same output path when they pass through a PBS3 in $\ket{R}/\ket{L}$ basis \cite{eisenberg2005multiphoton}, where $\ket{R}=1/\sqrt{2}\left(\ket{H}+i\ket{V}\right)$ and $\ket{L}=1/\sqrt{2}\left(\ket{H}-i\ket{V}\right)$. The PBS3 is constructed with a standard PBS in $\ket{H}/\ket{V}$ basis and four quarter-wave plates (QWPs), which transmits $\ket{R}$ state and reflects $\ket{L}$ one. We will demonstrate that the two bunching photons will not give a correct trigger signal to ruin the scheme in the main text. 

From Eq. (\ref{eq:outputState}) and Group 2, the state of the two bunching photon is in $\ket{H}_{2}\ket{V}_{2}$ or $\ket{+}_{3}\ket{-}_{3}$. Let's take the case $\ket{H}_{2}\ket{V}_{2}$ (which corresponds to case 0:2:0:2 in Group 2), and rewrite the state into $\ket{R}/\ket{L}$ basis:
\begin{align}
\ket{H}_{2}\ket{V}_{2}\nonumber
&=\frac{1}{\sqrt{2}}\left(\ket{R}_{2}+\ket{L}_{2}\right)\frac{-i}{\sqrt{2}}\left(\ket{R}_{2}-\ket{L}_{2}\right)\\\nonumber
&=\frac{-i}{2}\left(\ket{R}_{2}\ket{R}_{2}-\ket{R}_{2}\ket{L}_{2}+\ket{L}_{2}\ket{R}_{2}-\ket{L}_{2}\ket{L}_{2}\right)\\\nonumber
&\rightarrow\frac{1}{\sqrt{2}}\left(\ket{R}_{2}\ket{R}_{2}-\ket{L}_{2}\ket{L}_{2}\right)
\label{eq:photonbunchingeffect}
\end{align} 

After passing through the PBS3, the two photons in state $\ket{H}_{2}\ket{V}_{2}$ --- one in $\ket{H}$ and the other in $\ket{V}$ will go to the same output path of PBS3. Therefore, case 0:2:0:2 will not introduce any two-fold coincidence event between doctors at different output ports of PBS3. Similarly, we can get the same conclusion for the case 2:0:2:0 in Group 2, such that two photon states $\ket{+}_{3}\ket{-}_{3}$ --- one in $\ket{+}$ and the other in $\ket{-}$ will also go to the same output port of PBS3 in $\ket{R}/\ket{L}$ basis (because the state can be expanded into $1/\sqrt{2}(\ket{R}_{3}\ket{R}_{3}+\ket{L}_{3}\ket{L}_{3})$). Therefore, these cases will not introduce any two-fold coincidences.

For the cases in Group 3, the total photon number in path 2 and 3 does not equal to 2. When the totally photon number is zero (2:0:0:2) or one (1:1:0:2; 2:0:1:1), these cases will not give any coincidence detection events between ancillary photons. When the totally photon number is three (1:1:2:0; 0:2:1:1) or four (0:2:2:0), due to photons bunching effect as discussed in Group 2 (namely two photons are bunching into the same path), these cases could also contribute two-fold coincidence detection events between ancillary photons. However, with assistance of photon number resolving detectors (PNRDs) that can distinguish the number of photons \cite{fitch2003photon,nehra2020photon}, these cases will not generate a correct trigger signal.

Based on the calculation and discussions of the proposed scheme in Fig.1 in the main text, we can conclude that if there is a correct coincident count between ancillary photons, then 1-bit trigger information will be sent to do the related Pauli operations on the outputs, and then we will get the desired heralded CNOT gate. We also conclude that, for a high-performance heralded CNOT gate, a truly on-demand SPS together with PNRDs will be the key resources.

\hypertarget{part 02}{\subsection{2. Fidelity calculation for the heralded CNOT gate and event-ready entangled state}}

\hypertarget{part 02-1}{\subsection{(1). Fidelity calculation for the heralded CNOT gate}}

The fidelity calculation for our heralded CNOT operation is based on an efficient approach proposed by Hofmann \cite{hofmann2005complementary, hofmann2005quantum}. The fidelity is defined as the probability of obtaining the correct output averaged over all four possible inputs, as follows \cite{hofmann2005complementary}:
\begin{align}
F_{1}=&\frac{1}{4}\Big[P(HH|HH)+P(HV|HV)+P(VV|VH)+P(VH|VV)\Big],\\\nonumber
F_{2}=&\frac{1}{4}\Big[P(++|++)+P(--|+-)+P(-+|-+)+P(+-|--)\Big],
\label{eq:onlytwofidelity}
\end{align}
where $P(xy|mn)$ represents the output-input probabilities of the CNOT operation ($x$, $y$, $m$ and $n$ can be any polarization of $H$, $V$, $+$ and $-$).

To show the quantum behaviour of the CNOT gate, it was tested for different combinations of output-input states using complementary bases, that is, in both the computational basis ( $\ket{H}/\ket{V}$) and their linear superpositions basis ($\ket{+}/\ket{-}$). These eight measurement results described in Table.\ref{tab:fidelitycompute} (a) and (b) are already sufficient to obtain reliable estimates of the process fidelity $F_{1}$ and $F_{2}$ \cite{hofmann2005complementary}. We substitute the experimentally measured results into $F_{1}$ and $F_{2}$, and then we get $F_{1}=(87.8\pm2.1)\%$ in $\ket{H}/\ket{V}$ basis and $F_{2}=(88.6\pm2.1)\%$ in $\ket{+}/\ket{-}$ basis. These two complementary fidelities $F_{1}$ and $F_{2}$ define an upper and a lower bound for the quantum process fidelity $F_{proc}$ according to $\left(F_{1}+F_{2}-1\right)\leq F_{proc} \leq min\left(F_{1},F_{2}\right)$ \cite{hofmann2005complementary,bao2007optical}, which is ($76.4\pm2.9)\%\leq F_{proc} \leq(87.8\pm2.1$)\%.

Furthermore, we demonstrate that quantum parallelism has been achieved in our CNOT gate, providing that the gate cannot be reproduced by local operations and classical communications \cite{hofmann2005quantum}. Such quantum parallelism can be achieved if the average gate fidelity of three distinct conditional local operations exceeds 2/3, where $F_{1}$ and $F_{2}$ are just two of the required fidelities and the third required fidelity is  
\begin{align}
F_{3}=&\nonumber \frac{1}{4}\Big[P(RL|+H)+P(LR|+H)+P(RR|+V)+P(LL|+V)\\
&+P(RR|-H)+P(LL|-H)+P(RL|-V)+P(LR|-V)\Big].
\end{align}

We demonstrate it by preparing the control input in $\ket{+}/\ket{-}$ basis and target input in $\ket{H}/\ket{V}$ basis, and performing the measurement on the output state in $\ket{R}/\ket{L}$ basis. The experimentally measured result is also shown in Table.\ref{tab:fidelitycompute} (c), which gives fidelity $F_{3}=(87.0\pm2.2)\%$. The average gate fidelity of $F_{1}$, $F_{2}$ and $F_{3}$ is $(87.8\pm1.2)\%$, obviously exceeding the boundary of 2/3. 

\begin{table}[!htb]
	\begin{minipage}{.5\linewidth}
		\centering
		\begin{tabular}{c|c|c|c|c}
			out$\backslash$in& HH & HV & VH & VV \\
			\hline
			HH & 0.8884 & 0.1265 & 0.0108& 0.0082\\
			HV & 0.1095 & 0.8735 & 0.0194& 0.0123\\
			VH & 0.0020 & 0 & 0.1013& 0.8794\\
			VV & 0 & 0 & 0.8686& 0.1002\\	\hline
		\end{tabular}
		\\(a) In $\ket{H}/\ket{V}$ basis
    	\label{tab:computationbasis}
	\end{minipage}%
	\begin{minipage}{.6\linewidth}
		\centering
		\begin{tabular}{c|c|c|c|c}
			out$\backslash$in& ++ & +- & -+ & -- \\
			\hline
			++ & 0.8684 & 0.0102 & 0.0575& 0.0265\\
			+- & 0 & 0.0832 & 0.004& 0.8458\\
			-+ & 0.1299 & 0.0153 & 0.9385& 0.0145\\
			-- & 0.0018 & 0.8913 & 0& 0.1133\\	\hline
		\end{tabular}
	\\(b) In $\ket{+}/\ket{-}$ basis
	\label{tab:superpositionbasis}
	\end{minipage} 
	\begin{minipage}{.9\linewidth}
	\centering
		\begin{tabular}{c|c|c|c|c}
			out$\backslash$in& +H & +V & -H & -V \\
			\hline
			RR & 0.0751 & 0.3921 & 0.4120& 0.0531\\
			RL & 0.3529 & 0.0608 & 0.0469& 0.4134\\
			LR & 0.4970 & 0.0578 & 0.0996& 0.4804\\
			LL & 0.0751 & 0.4894 & 0.4336& 0.0531\\	\hline
		\end{tabular}
	\\ (c) In different basis
	\label{tab:differentbasis}
    \end{minipage} 
   \caption{The experimentally achieved output-input probabilities of our CNOT gate.}
	\label{tab:fidelitycompute}
\end{table}


In the real experiment, we achieve the heralded CNOT gate with a rate of ~85 per minute, confirmed by detecting 4-fold coincidence count rates when the heralded signals are correct. Note that, in order to have a good balance between the gate fidelity and the operation rates of the heralded CNOT gates, we reduce the excitation pulse power to increase the purity and indistinguishability of the produced single photons (these are the main factors for getting a high-performance heralded CNOT gate). When one increase the excitation pulse power to $\pi$ pulse at $\sim$76MHz pump rate, the purity and indistinguishability will decrease to $\sim$0:06(1) and $\sim$0.86(1) at 6.5$\mu s$ time delay (all values are measured without any filters and corrections). However, the count rate of resonance florescence single photons will increase to $\sim20$ MHz at the single-mode fiber output end of optical confocal system (detected by superconductor nanowire single photon detector (SNSPD) with a detection efficiency of $\sim 80\%$). The fidelity of the CNOT gate will drop to a range from 0.636 to 0.740 (this value is estimated with the direct measurement value of SPS's indistinguishability), however the operation rate of heralded CNOT gate will increase to $\sim 3320$ per minute.

\hypertarget{part 02-2}{\subsection{(2). Fidelity calculation for the event-ready Bell state}}

The fidelity of an entangled photon state is defined as the overlap of experimentally generated state and the theoretical ideal one \cite{guhne2009entanglement, wang2016experimental,li2020multiphoton}:
$F(\psi^{N})=\bra{\psi^{N}}\rho_{expe}\ket{\psi^{N}}=\operatorname{Tr}(\rho_{\text{ideal}} \rho_{\text {expe}})=\left\langle\ket{\psi^{N}}\bra{\psi^{N}}\right\rangle$. $\ket{\psi^{N}}$ represents the ideal $N$-photon state, $\rho_{\text{ideal}}$ ($\rho_{\text {expe}}$) denotes the density matrix of ideal state (experimentally generated state) and $\rho_{\text{ideal}}=\ket{\psi^{N}}\bra{\psi^{N}}$. For the Greenberger-Horne-Zeilinger (GHZ) state, $F\left(\psi^{N}\right)=\left(\left\langle P^{N}\right\rangle+\left\langle C^{N}\right\rangle\right) / 2$, with $\left\langle P^{N}\right\rangle=\left\langle(|H\rangle\langle H|)^{\otimes N}\right\rangle+\left\langle(|V\rangle\langle V|)^{\otimes N}\right\rangle$, which denotes the population of $|H\rangle^{\otimes N}$ and $|V\rangle^{\otimes N}$ components of the generated GHZ state. In addition, 
\begin{equation}
\left\langle C^{N}\right\rangle\!=\!\frac{1}{N}\!\sum_{k=0}^{N-1}(-1)^{k}\!\left\langle\!M_{k \pi / N}^{\otimes N}\right\rangle,
\end{equation} 
with $M_{k\pi/N}=\sigma_{x}\cos(k\pi/N)+\sigma_{y}\sin(k\pi/N)$, where $\sigma_{x}$ and $\sigma_{y}$ are Pauli operators, and $k=0, 1, ..., N-1$. The $\left\langle C^{N}\right\rangle$ is defined by the off-diagonal elements of the density matrix and reflects the population of coherent superposition between the $|H\rangle^{\otimes N}$ and $|V\rangle^{\otimes N}$ components of the generated GHZ state. The $M_{k\pi/N}$ is a single qubit observable with eigenstate of $|\psi\rangle^{\pm k\pi/N}=\left(|H\rangle \pm e^{i k\pi/N}|V\rangle\right) / \sqrt{2}$  and eigenvalue of $\pm1$. The single qubit projective measurement can be realized by a pair of wave plates and a PBS.

For the Bell state $\ket{\Phi^{+}}_{14}=1/\sqrt(2)(\ket{H}_{1}\ket{H}_{4}+\ket{V}_{1}\ket{V}_{4})$, it is equivalent to a GHZ state $\psi^{N}$ when $N=2$. So its fidelity can be measured followed by the standard method of GHZ state (the above description). However, in our experiment, we produce the entangled photon pair state in 
$\ket{\Psi^{-}}_{14}=1/\sqrt(2)(\ket{H}_{1}\ket{V}_{4}-\ket{V}_{1}\ket{H}_{4})=\sigma _{z1}\sigma _{x4}\ket{\Phi^{+}}_{14}$, where $\sigma_{z1}$ and $\sigma_{x4}$ are the Pauli operators operated on photons in path 1 and 4. So the fidelity of $F_{\ket{\Psi^{-}}_{14}}$ will be:
\begin{align}
{F_{{{\left| {{\Psi^ - }} \right\rangle }_{14}}}}   \nonumber
&= Tr({\rho _{\exp e}}{\left| {{\Psi ^ - }} \right\rangle _{14}}\left\langle {{\Psi ^ - }} \right|) = \left\langle {{{\left| {{\Psi ^ - }} \right\rangle }_{14}}\left\langle {{\Psi ^ - }} \right|} \right\rangle  \\ \nonumber
&= \left\langle {{\sigma _{z1}}{\sigma _{x4}}{{\left| {{\Phi ^ + }} \right\rangle }_{14}}\left\langle {{\Phi ^ + }} \right|{\sigma _{z1}}{\sigma _{x4}}} \right\rangle  \\ \nonumber
&= \frac{1}{2}\Big(\left\langle {{\sigma _{z1}}{\sigma _{x4}}{P^{2}}{\sigma _{z1}}{\sigma _{x4}}} \right\rangle  + \left\langle {{\sigma _{z1}}{\sigma _{x4}}{C^{2}}{\sigma _{z1}}{\sigma _{x4}}} \right\rangle \Big) \\ 
&= \frac{1}{2}\Big(\left\langle {P^{2-}} \right\rangle  + \left\langle {C^{2-} } \right\rangle \Big)
\label{eq:fidelityforpsai}
\end{align}
here the $P^{2}={\left| {HH} \right\rangle _{14}}\left\langle {HH} \right| + {\left| {VV} \right\rangle _{14}}\left\langle {VV} \right|$ and $C^{2}={\frac{1}{2}(M_{{{0\pi } \mathord{\left/{\vphantom {{0\pi } 2}} \right.\kern-\nulldelimiterspace} 2}}^{ \otimes 2} - M_{{\pi \mathord{\left/{\vphantom {\pi  2}} \right.\kern-\nulldelimiterspace} 2}}^{ \otimes 2})}= {\frac{1}{2}({\sigma _{x1}} \otimes {\sigma _{x4}} - {\sigma _{y1}} \otimes {\sigma _{y4}})}$. Then we will get: 
\begin{align}
\left\langle P^{2-}\right\rangle\nonumber
&= \left\langle{\sigma _{z1}}{\sigma _{x4}}\Big({\left| {HH} \right\rangle _{14}}\left\langle {HH} \right| + {\left| {VV} \right\rangle _{14}}\left\langle {VV} \right|\Big){\sigma _{z1}}{\sigma _{x4}}\right\rangle\\  
&= \left\langle\Big({\left| {HV} \right\rangle _{14}}\left\langle {HV} \right| + {\left| {VH} \right\rangle _{14}}\left\langle {VH} \right|\Big) \right\rangle \label{eq:noncoherentitempsai} \\ \nonumber
\left\langle C^{2-} \right\rangle\nonumber
&= \left\langle{\sigma _{z1}}{\sigma _{x4}}\left[ {\frac{1}{2}({\sigma _{x1}} \otimes {\sigma _{x4}} - {\sigma _{y1}} \otimes {\sigma _{y4}})} \right]{\sigma _{z1}}{\sigma _{x4}}\right\rangle \\ 
&= \left\langle - \frac{1}{2}({\sigma _{x1}} \otimes {\sigma _{x4}} + {\sigma _{y1}} \otimes {\sigma _{y4}})\right\rangle
\label{eq:coherentitempsai}
\end{align}

For measuring the average value of $\left\langle P^{2-}\right\rangle$, one just need to detect the generated state in $\ket{H}/\ket{V}$ basis and compute the probability of items $\ket{HV}_{14}$ and $\ket{VH}_{14}$. For measuring the average value of $\left\langle C^{2-}\right\rangle$, one needs to detect the generated state in $\ket{+}/\ket{-}$ and $\ket{R}/\ket{L}$ basis for the average value of items $\left\langle\sigma _{x1}\otimes \sigma _{x4}\right\rangle$ and $\left\langle\sigma _{y1}\otimes \sigma _{y4}\right\rangle$. In the real experiment, the values of $\left\langle {P_2^ - } \right\rangle$, $\left\langle\sigma _{x1}\otimes \sigma _{x4}\right\rangle$ and $\left\langle\sigma _{y1}\otimes \sigma _{y4}\right\rangle$ are 0.8729, -0.8039, and -0.7875, indicating the fidelity of the heralded Bell state $\ket{\Psi^{-}}_{14}$ is $83.4\pm2.4\%$.

\hypertarget{part 03}{\subsection{3. Detailed analysis of the heralding efficiency}}

Our heralded CNOT operation depends on the two ancillary photons' measurement outcome and their corresponding Pauli operators, as shown in Fig.1 of main text. For an ideal heralded CNOT gate, perfect on-demand single-photon sources (SPSs) with unity efficiency and ideal photon-number-resolving detectors (PNRDs) which can completely distinguish the number of photons will be the key resources. However, due to the imperfect SPS and PNRDs, there are some cases which will happen and contribute heralded signal while the desired CNOT operation does not occur (please refer to \hyperlink{part 01-2}{section 1.(2)}). Such a performance can be identified by heralding efficiency, a critical parameter for large-scale quantum information processing. Generally speaking, the heralding efficiency $\eta_{h}$ is, for a given input state, the probability to obtain the expected CNOT gate operation when heralded signals are given by performing the Bell states measurement on the two ancillary photons. For a perfect scheme (all implementations and elements in the experiments are ideal), the heralding efficiency should be unity.

For simplicity, we define experimental heralding efficiency $\eta_{h}$, which is given by the ratio between the probability of four-fold coincidence detection $P_{4}$ (desired CNOT operations that are detected) and the probability of two-fold coincidence detection $P_{2}$ (heralding signals). Therefore, the heralding efficiency $\eta_{h}$ is described as
\begin{equation}
\eta_{h}=P_{4}/P_{2}.
\end{equation}

In our experiment, we use four superconductor nanowire single-photon detectors (SNSPDs) without any photon-number-resolving ability, to construct each pseudo-PNRD (as shown in Fig.2 of main text). We define $\eta_d$ as the detection efficiency of a single SNSPD (we assume that every SNSPD has the same value $\eta_{d}\sim$ 0.8 in the experiment), thus the probability to distinguish two-photon events is only $3\eta_{d}^2/4$ for the pseudo-PNRD. That means the pseudo-PNRDs can only exclude a part of unwanted coincidence detection events between heralded detectors $D_{1}$ ($D_{H1}$ or $D_{V1}$) and $D_{2}$ ($D_{H2}$ or $D_{V2}$). Thus the heralding efficiency $\eta_{h}$ will be between the ideal value obtained by using truly PNRDs and the value given by using standard detectors (which can only distinguish zero and non-zero photon cases). Here, we calculate the experimental heralding efficiencies for implementations using pseudo-PNRDs with arbitrary two-photon resolvable capability (in our scheme, we just need to distinguish one and two photons for the heralding detection and the photon number resolvable capability can be tuned by changing $\eta_d$). We define $P_{(m|n|k)}$ as the probability of detecting only $m$ photons when inject $n$ photons to a $k$-photon resolvable pseudo-PNRD ($m$, $n$ and $k$ are integers and $m\leqslant n$) \cite{fitch2003photon, nehra2020photon}. By detecting any n-photon pulse, there should be $\sum_{i=0}^{n}P_{(i|n|k)}=1$ and $P_{(i|n|k)}=0$ when $i> k$. Thus, when a two-photon pulse inject to a four-photon resolvable pseudo-PNRDs (our pseudo-PNRD), the probabilities of detecting two photons, only one photon and zero photon are:
\begin{align}
P_{(2|2|4)}=\frac{3}{4}\eta_{d}^2,\quad P_{(1|2|4)}=\frac{3}{4}\eta_{d}(1-\eta_{d})*2+\frac{1}{4}\eta_{d},\quad P_{(0|2|4)}=\frac{3}{4}(1-\eta_{d})^2+\frac{1}{4}(1-\eta_{d})
\label{eq:pseudo-PNRD}
\end{align}
respectively, satisfying $P_{(2|2|4)}+P_{(1|2|4)}+P_{(0|2|4)}=1$. When only one-photon pulse enters into our pseudo-PNRDs, we have $P_{(1|1|4)}=\eta_{d}$ and $P_{(0|1|4)}=1-\eta_{d}$. In addition, for an ideal PNRD, $P^{ide}_{(2|2|4)}=1$ and others are zero. For a standard photon detector, which just can distinguish zero and non-zero photons, when $n$ photons enter to the detector, one can have $P^{sta}_{(1|n|1)}=\eta_{d}$ and $P^{sta}_{(0|n|1)}=1-\eta_{d}$, and all others are zero.

Then, we assume that the efficiency of photon brightness at the fiber output of confocal system is $\eta_{f}$, optical switch (means demultiplexer) efficiency is $\eta_{w}$, optical line efficiency is $\eta_{l}$ and all SNSPDs has the same detection efficiency $\eta_{d}$. For simplicity, we suppose that all losses of demultiplexer and optical lines in the experiment are uniform such that one could instead use a total single photon efficiency $\eta_{s}=\eta_{f}\eta_{w}\eta_{l}$ as the overall efficiency of every one photon streams at the CNOT inputs. The heralding efficiency $\eta_{h}$ is now related to the $\eta_{s}$, and then the mixed state of the source (without taking the purity of SPSs into account) can be assumed as
\begin{equation}
\rho=(1-\eta_{s})\ket{0}\bra{0}+\eta_{s}\ket{1}\bra{1}
\end{equation}
at the CNOT inputs, where $\ket{0}$ and $\ket{1}$ stand for vacuum Fock state and the single photon state, respectively.   

Now we calculate the heralding efficiency $\eta_{h}$, which is given by $\eta_{h}=P_{4}/P_{2}$. Following the scheme in \hyperlink{part 01-1}{section 1.(1)}, the successful probability of having four-fold coincidence detection is $1/8$ \cite{bao2007optical}. Therefore, we have:
\begin{equation}
P_{4}=\eta_{s}^{4}*P_{(1|1|4)}^{4}/8
\end{equation}
for correct heralded CNOT gates (each pseudo-PNRD detects only one photon). 

We note that $P_{2}$ depends on specific polarization states in the CNOT inputs, particularly when there are less than four photons entering into the CNOT gate. Different combinations of input photons at the input port of heralded CNOT gate will lead to different $P_{2}$. We set the input states as the general states which are $\ket{\psi}_{c_{in}}=\alpha\ket{H}+\beta\ket{V}$ and $\ket{\psi}_{t_{in}}=\gamma\ket{H}+\delta\ket{V}$, and two ancillary photons are $\ket{\psi}_{a_{1}}=1/\sqrt{2}\left(\ket{H}+\ket{V}\right)$, $\ket{\psi}_{a_{2}}=\ket{H}$ as described in \hyperlink{part o1}{section 1}.

First, we consider the situation of using pseudo-PNRDs and study different cases for computing $P_{2}$ as follows:

(1). Four single photons are produced at the CNOT inputs and there are correct two-fold coincidence events for the heralded CNOT gate. Such a case is described in Group 1 of \hyperlink{part 01-2}{section 1.(2)}, which means $1:1:1:1$ photon distribution in path 1, 2, 3 and 4. Hence, the probability for $P_{2}$ in this case (defined as $P_{2-4-2}$) equals to the probability $P_{4}$, which is 
\begin{equation}
P_{2-4-2}=P_{4}=\eta_{s}^{4}*P_{(1|1|4)}^{4}/8,
\label{eq:p-2-4-2}
\end{equation} 
where $P_{2-4-2}$ means the probability that detecting a two-fold coincidence event (the first label: 2) when four photons enter into the heralded CNOT gate (the second label: 4) and total two photons (the third label: 2) are in path 2 and 3 (the meaning of the similar form as $P_{2-4-2}$ holds for the rest of the text). 

For the case 1:1:2:0 (or 0:2:1:1) in Group 3 described in \hyperlink{part 01-2}{section 1.(2)}, the states of photons in path 2 and 3 are just corresponding to items $\ket{H}_{2}\ket{+}_{3}\ket{-}_{3}$ and $\ket{V}_{2}\ket{+}_{3}\ket{-}_{3}$ (or $\ket{H}_{2}\ket{V}_{2}\ket{+}_{3}$ and $\ket{H}_{2}\ket{V}_{2}\ket{-}_{3}$). For the state of $\ket{H}_{2}\ket{+}_{3}\ket{-}_{3}$, it will be produced with a probability of $\eta_{s}^4\left |\alpha(\gamma+\delta)\right |^2/8$ (please refer to Eq.(\ref{eq:outputState})). There totally are three photons in path 2 and 3, however, with the help of photon bunching effect as described in \hyperlink{part 01-2}{section 1.(2)}, two photons in path 3 will all in states either $\ket{R}_3$ or $\ket{L}_3$ and go to the same output of PBS3 in $\ket{R}/\ket{L}$ basis. This item will have a probability of $1/2$ for photons output from different ports of PBS3. Then we consider the item states that can contribute a heralded signal, if two bunching photons go together to one pseudo-PRND $D_{H}$ or $D_{V}$ that just detected one photon signal, the probability is $P_{(1|2|4)}*P_{(1|1|4)}/2$; if the two bunching photons go to different pseudo-PRNDs $D_{H}$ and $D_{V}$ that just detected one photon signal, the probability is $P_{(0|1|4)}*P_{(1|1|4)}*P_{(1|1|4)}$. Therefore, this item will give a heralding signal with a probability:
\begin{align}
&\eta_{s}^4*\frac{\left |\alpha(\gamma+\delta)\right |^2}{8}*\frac{1}{2}*(\frac{P_{(1|2|4)}P_{(1|1|4)}}{2}+P_{(0|1|4)}P_{(1|1|4)}P_{(1|1|4)}) \\ \nonumber
&=\frac{\eta_{s}^4\left |\alpha(\gamma+\delta)\right |^2}{32}(P_{(1|2|4)}*P_{(1|1|4)}+2P_{(0|1|4)}*P_{(1|1|4)}^2)
\end{align}
and then all the case 1:1:2:0 and 0:2:1:1 will contribute a probability:
\begin{align}
P_{2-4-3} \nonumber
&=\frac{\eta_{s}^4}{32}\Big(P_{(1|2|4)}P_{(1|1|4)}+2P_{(0|1|4)}P_{(1|1|4)}^2\Big)\Big[\left |\alpha(\gamma+\delta)\right |^2+\left |\beta(\gamma+\delta)\right |^2+\left |\alpha(\gamma+\delta)\right |^2+\left |\alpha(\gamma-\delta)\right |^2\Big] \\
&=\frac{\eta_{s}^4}{32}\Big(P_{(1|2|4)}P_{(1|1|4)}+2P_{(0|1|4)}P_{(1|1|4)}^2\Big)\Big(\left |\gamma+\delta\right |^2+2\left |\alpha\right |^2\Big)	
\label{eq:p-2-4-3}
\end{align}

For the case 0:2:2:0 in group 3 described in \hyperlink{part 01-2}{section 1.(2)}, it will contribute to a probability of $\eta_{s}^4\left |\alpha(\gamma+\delta)\right |^2/8$. Also with the help of photon bunching effect, there is $1/2$ probability for photons output from different ports of PBS3 in $\ket{R}/\ket{L}$ basis. Then detected two-fold coincidence detection will give a probability of $P_{(1|2|4)}^2/4+P_{(1|2|4)}*P_{(0|1|4)}*P_{(1|1|4)}+P_{(0|1|4)}^2*P_{(1|1|4)}^2=\Big(P_{(1|2|4)}/2+P_{(0|1|4)}P_{(1|1|4)}\Big)^2$. Thus this case will contribute two-fold heralded signal with a probability:
\begin{align}
P_{2-4-4}   \nonumber
&=\eta_{s}^4*\frac{\left |\alpha(\gamma+\delta)\right |^2}{8}*\frac{1}{2}*\Big(\frac{P_{(1|2|4)}}{2}+P_{(0|1|4)}*P_{(1|1|4)}\Big)^2 \\
&=\frac{\eta_{s}^4}{64}\Big(P_{(1|2|4)}+2P_{(0|1|4)}P_{(1|1|4)}\Big)^2\left|\alpha(\gamma+\delta)\right |^2
\label{eq:p-2-4-4}
\end{align}

Thus the total probability of detecting two-fold coincidence in this situation is: $$P_{2-4}=P_{2-4-2}+P_{2-4-3}+P_{2-4-4}.$$

(2). Only three single photons are produced at the heralded CNOT inputs and there are two-fold coincidence events for heralding signal. There are four possible combinations for three single photons at the CNOT inputs (i.e. $\tbinom{4}{3}=4$), and they will contribute to the similar probability formula for heralding signals. For example, the mode combination of input photons such as $(a_{1}, a_{2}, t_{in})$, there are two items 0:1:1:1 and 0:1:2:0 that will give contributions.

For the item 0:1:1:1 of the case $(a_{1}, a_{2}, t_{in})$, this is equivalent to the quantum teleportation. Our Bell state measurement can only distinguish two of four Bell states, in the end we have the probability for the heralding signal is:
\begin{equation}
\eta_{s}^3(1-\eta_{s})*\frac{1}{2}*\frac{1}{2}*\frac{1}{2}P_{(1|1|4)}^2=\frac{\eta_{s}^3(1-\eta_{s})}{8}P_{(1|1|4)}^2
\end{equation} 

So the probability of contributing to the heralding signal under the case there are total two photons in path 2 and 3 is:
\begin{align}
P_{2-3-2} 
&=\frac{\eta_{s}^3(1-\eta_{s})P_{(1|1|4)}^2}{2} \Big(\frac{\left |\alpha\right |^2}{2}+\frac{1}{4}+\frac{\left |\gamma+\delta\right |^2}{4}+\frac{1}{4}\Big) 
\label{eq:p-2-3-2-01} \\
&=\frac{\eta_{s}^3(1-\eta_{s})P_{(1|1|4)}^2}{8} \Big(2\left |\alpha\right |^2+\left |\gamma+\delta\right |^2+2\Big)
\label{eq:p-2-3-2}
\end{align} 
Items in the brackets of Eq.(\ref{eq:p-2-3-2-01}) is related to the input case $(c_{in}, a_{2}, t_{in})$, $(a_{1}, a_{2}, t_{in})$, $(c_{in}, a_{1}, t_{in})$ and $(c_{in}, a_{1}, a_{2})$.

For the item 0:1:2:0 of the case $(a_{1}, a_{2}, t_{in})$, its probability is partly similar to the item 1:1:2:0, since they have the same state of photons entering to path 2 and 3. The calculated probability is
\begin{align}
&\nonumber \eta_{s}^3(1-\eta_{s})*\frac{1}{2}*\frac{\left|\gamma+\delta\right |^2}{4}*\frac{1}{2}\Big(\frac{P_{(1|2|4)}*P_{(1|1|4)}}{2}+P_{(0|1|4)}*P_{(1|1|4)}*P_{(1|1|4)}\Big) \\
&=\frac{\eta_{s}^3(1-\eta_{s})\left |\gamma+\delta\right |^2}{32}\Big(P_{(1|2|4)}*P_{(1|1|4)}+2P_{(0|1|4)}*P_{(1|1|4)}^2\Big)
\end{align}

Therefore, the probability for detecting heralding signal with all three photons in path 2 and 3 is:
\begin{align}
P_{2-3-3} \nonumber
&=\frac{\eta_{s}^3(1-\eta_{s})}{4}\Big(P_{(1|2|4)}P_{(1|1|4)}+2P_{(0|1|4)}P_{(1|1|4)}^2\Big)\Big[\frac{\left |\alpha(\gamma+\delta)\right |^2}{4}+\frac{\left |\gamma+\delta\right|^2}{8}+\frac{\left |\alpha(\gamma+\delta)\right |^2}{4}+\frac{\left |\alpha\right |^2}{4}\Big]  \\
&=\frac{\eta_{s}^3(1-\eta_{s})}{32}\Big(P_{(1|2|4)}P_{(1|1|4)}+2P_{(0|1|4)}P_{(1|1|4)}^2\Big)\Big[(4\left |\alpha\right |^2+1)\left |\gamma+\delta\right |^2+2\left |\alpha\right |^2\Big]
\label{eq:p-2-3-3}
\end{align} Items in the brackets of Eq.(\ref{eq:p-2-3-3}) is related to the input case $(c_{in}, a_{2}, t_{in})$, $(a_{1}, a_{2}, t_{in})$, $(c_{in}, a_{1}, t_{in})$ and $(c_{in}, a_{1}, a_{2})$. In the end, the total probability of three photons at the inputs of a heralded CNOT gate is $$P_{2-3}=P_{2-3-2}+P_{2-3-3}.$$

(3). Only two single photons are produced at the CNOT inputs and there are two-fold coincidence events. Due to the photon bunching effect, there are only four possible combinations that can contribute two-fold coincidence events for heralded detectors. One photon input is from $c_{in}$ or $a_{1}$ and another photon input from $a_{2}$ or $t_{in}$. After they pass through PBS1 and PBS2, they have the same probability for having two-fold coincidence events. Let's say that the two photons are in modes $c_{in}$ and $t_{in}$, they will go to path 2 and 3 with a probability of $\left |\alpha(\gamma+\delta)\right |^2/2$, and will have $P_{(1|1|4)}^2/2$ probability to a two-fold coincidence. Thus, this case will contribute to heralding signal with a probability $\eta_{s}^2(1-\eta_{s})^2\left |\alpha(\gamma+\delta)\right |^2P_{(1|1|4)}^2/4$. The total probability for all the four cases will be:
\begin{align}
P_{2-2}    \nonumber
&=\eta_{s}^2(1-\eta_{s})^2\frac{P_{(1|1|4)}^2}{2}\Big[\frac{\left |\alpha(\gamma+\delta)\right |^2}{2}+\frac{\left |\alpha\right |^2}{2}+\frac{1}{4}+\frac{\left |\gamma+\delta\right |^2}{4}\Big] \\ 
&=\eta_{s}^2(1-\eta_{s})^2\frac{P_{(1|1|4)}^2}{8}\Big[2 \left |\alpha(\gamma+\delta)\right |^2+2\left |\alpha\right |^2+\left |\gamma+\delta\right |^2+1\Big]
\label{eq:p-2-2}
\end{align} 

Finally, we can get the probability $P_{2}$ of two-fold coincidence events for heralded signal as $P_{2}=P_{2-2}+P_{2-3}+P_{2-4}$, leading to the heralded efficiency:
\begin{equation}
\eta_{h}=\frac{P_{4}}{P_{2}}=\frac{P_{4}}{P_{2-2}+P_{2-3-2}+P_{2-3-3}+P_{2-4-2}+P_{2-4-3}+P_{2-4-4}}
\label{eq:finalefficiency-01}
\end{equation}

Then we substitute Eqs.(\ref{eq:p-2-2}), (\ref{eq:p-2-3-2}), (\ref{eq:p-2-3-3}), (\ref{eq:p-2-4-2}), (\ref{eq:p-2-4-3}), (\ref{eq:p-2-4-4}) into Eq.(\ref{eq:finalefficiency-01}), we will get the finally formula of $\eta_{h}$, which is a function of $\alpha$, $\beta$, $\gamma$, $\delta$, $\eta_{s}$, $\eta_{d}$. For simplicity, we set $\alpha=\beta=1/\sqrt{2}$, $\gamma=1, \delta=0$, which means the control and target qubits are $\ket{\psi}_{c_{in}}\ket{\psi}_{t_{in}}=\ket{+}_{c_{in}}\ket{H}_{t_{in}}$. After the CNOT gate operation, we will get a maximally entangled Bell state. With the above parametric setting, we have the following equations: 
\begin{align}
&P_{4}=\eta_{s}^{4}P_{(1|1|4)}^{4}/8;\quad \quad \quad \quad \quad \quad \quad P_{2-2}=\eta_{s}^2(1-\eta_{s})^2P_{(1|1|4)}^2/2;\\
&P_{2-3-2}=\eta_{s}^3(1-\eta_{s})P_{(1|1|4)}^2/2;\quad  P_{2-3-3}=\eta_{s}^3(1-\eta_{s})\Big(P_{(1|2|4)}*P_{(1|1|4)}+2P_{(0|1|4)}*P_{(1|1|4)}^2\Big)/8;\\ &P_{2-4-2}=\eta_{s}^{4}P_{(1|1|4)}^{2}/8;\quad \quad \quad \quad P_{2-4-3}=\eta_{s}^4\Big(P_{(1|2|4)}*P_{(1|1|4)}+2P_{(0|1|4)}*P_{(1|1|4)}^2\Big)/16; \\
&P_{2-4-4}=\eta_{s}^4 \Big(P_{(1|2|4)}+2P_{(0|1|4)}*P_{(1|1|4)}\Big)^2/128;
\label{eq:P-2-4-4-02}
\end{align}

Now we bring the real detector's efficiency $\eta_{d}\approx 0.8$ into Eq:(\ref{eq:pseudo-PNRD}), yielding $P_{(1|2|4)}=0.44$, $P_{(0|2|4)}=0.08$, $P_{(1|1|4)}=0.8$, $P_{(0|1|4)}=0.2$. Then we use these parameters for Eq.(\ref{eq:finalefficiency-01}-\ref{eq:P-2-4-4-02}), giving the heralded efficiency $\eta_{h}^{pse}$:
\begin{equation}
\eta_{h}^{pse}=\frac{\eta_{s}^2}{(2-0.7625\eta_{s})^2}
\label{eq:heraldedefficiency-pseudo-PNRD}
\end{equation} 
where the superscript stands for the situation of using pseudo-PNRDs. 

In the real experiment, under $\pi$ pulse resonant excitation with a repetition rate $\sim$76 MHz, $\sim$16 MHz polarized resonance fluorescence single photons could be maximally registered by a single-mode fiber coupled SNSPD (detector efficiency $\sim$80\%). We now take these real experimental parameters $\eta_{f}\approx 16M/0.8/76M=0.263$, $\eta_{w}\approx0.83$, $\eta_{l}\approx0.80$ and $\eta_{det}\approx0.80$ ($\eta_{s}=\eta_{f}\eta_{w}\eta_{l}\approx0.175$) into the Eq.(\ref{eq:heraldedefficiency-pseudo-PNRD}), and now we will obtain an experimental heralding efficiency $\eta_{h}^{pse}=0.00875$.

In the case of an ideal two-photon resolvable PNRDs (that can completely distinguish the number of photons), we have $P_{(2|2|4)}=1$, $P_{(1|2|4)}=0$, $P_{(0|2|4)}=0$, $P_{(1|1|4)}=1$ and $P_{(0|1|4)}=0$. We bring these parameters into Eq.(\ref{eq:finalefficiency-01}-\ref{eq:P-2-4-4-02}), which gives us the heralding efficiency $\eta_{h}^{ide}$:
\begin{equation}
\eta_{h}^{ide}=\frac{\eta_{s}^2}{(2-\eta_{s})^2}
\label{eq:heraldedefficiency-PNRD}
\end{equation}

In the case of the standard photon detectors (that can distinguish zero and non-zero photons), we have $P_{(2|2|4)}=0$, $P_{(1|2|4)}=P_{(1|1|4)}=\eta_{d}=0.8$ and $P_{(0|1|4)}=P_{(0|2|4)}=1-\eta_{d}=0.2$. We also substitute them into Eq.(\ref{eq:finalefficiency-01}-\ref{eq:P-2-4-4-02}), which gives us the heralding efficiency $\eta_{h}^{sta}$:
\begin{equation}
\eta_{h}^{sta}=\frac{\eta_{s}^2}{(2-0.65\eta_{s})^2}
\label{eq:heraldedefficiency-non-PNRD}
\end{equation}

\begin{figure}[!h]
	\includegraphics[width=0.66\textwidth]{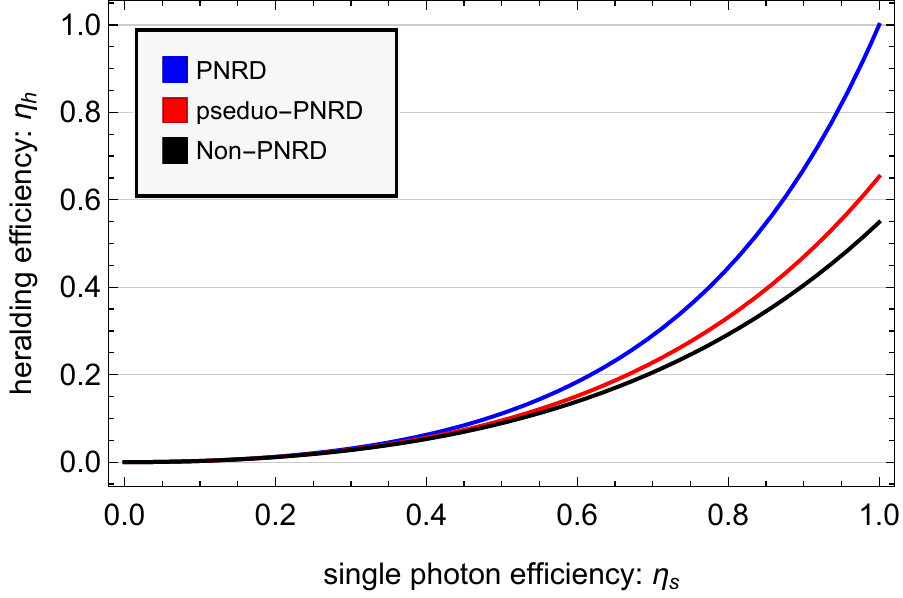}
	\caption{For a given input states $\ket{+}_{c}\ket{H}_{t}$ and detector efficiency $\eta_{d}=0.8$, the heralding efficiency $\eta_{h}$ varies with the single photon efficiency $\eta_{s}$ for implementations using PNRDs, pseudo-PNRDs and Non-PNRDs.}  
	\label{fig:heralding}
\end{figure}

Furthermore, we take the experimental number $\eta_{s}\approx 0.175$ into Eqs. (\ref{eq:heraldedefficiency-non-PNRD}) and (\ref{eq:heraldedefficiency-PNRD}) and we will get $\eta_{h}^{sta}=0.00857$, $\eta_{h}^{ide}=0.00915$. When one varies the total single photon efficiency $\eta_{s}$, the related heralding efficiency $\eta_{h}$ will increase (we show their changes in Fig.\ref{fig:heralding}). Although the numbers of $\eta_{h}^{ide}$, $\eta_{h}^{pse}$ and $\eta_{h}^{sta}$ are similar in current experimental setting (mainly limited by the relatively low $\eta_{s}$), one can further improve the $\eta_{h}^{pse}$ to near-unity by increasing $\eta_{d}$ and $\eta_{s}$ close to one. For example, one might achieve near-unity $\eta_{s}$ by coupling QDs to an asymmetric micro-cavity \cite{wang2019towards} and increasing demultiplexer efficiency. Additionally, the indistinguishability and purity of QD-based SPSs can be both improved to near-unity \cite{senellart2017high,he2013demand}. Therefore, with the assistance of pseudo-PNRD and a high-quality SPSs, one could obtain a heralded CNOT gate with high fidelity, high count rates and high heralding efficiency. 

\hypertarget{part 04}{\subsection{4. The details of the demultiplexer}}
\begin{figure}[!b]
	\includegraphics[width=0.9\textwidth]{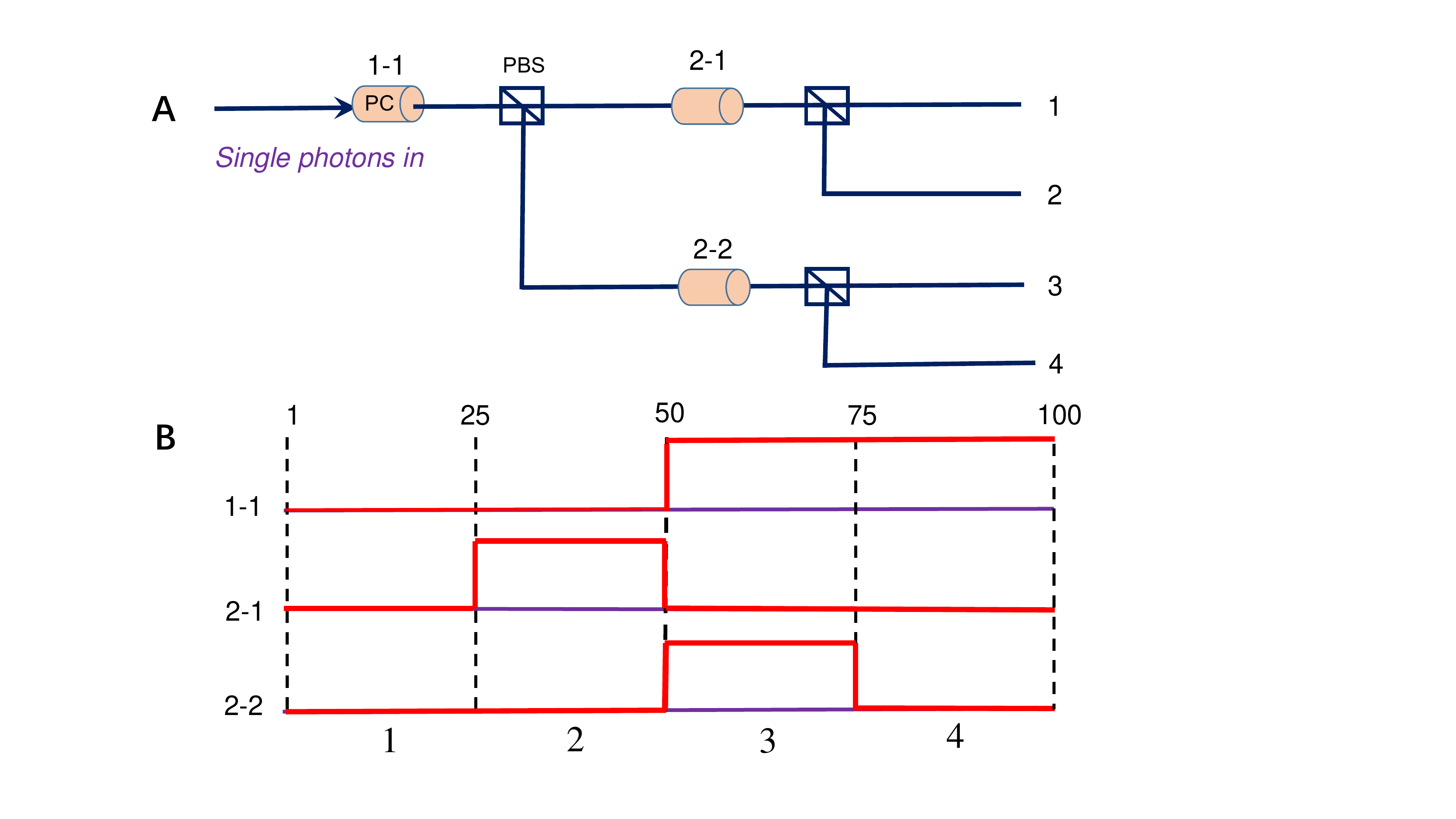}
	\caption{The architecture of the demultiplexer used in our experiment. \textbf{A}. A stream of single photon pulses are actively evenly divided into four free spatial modes by the demultiplexer consisted of 3 pairs of Pockels cells (PCs) and PBSs. \textbf{B}. Electric signals for the corresponding PCs in \textbf{A}. The width and delay for all signals must be well controlled to achieve correct switching operation.}
	\label{fig:switches}
\end{figure} 

Our 4-channel demultiplexer consists of 3 pairs of Pockels cells (PCs) and polarized beam splitters (PBSs) (see the Fig.2 in the main text and Fig.\ref{fig:switches} (A)), which are both customized and anti-reflective. Each PC is used to actively change the photon polarization by 90$^{\circ}$ with $\sim1800$ volts, which is driven by a high-voltage driver and a function generator. All PCs are synchronized to the pulse laser and a coincidence count unit \cite{wang2017high}. They are operated at a repetition rate of 0.76 MHz, which is 1/100 of the laser repetition rate 76 MHz. In this case, every 100 photon pulses are evenly divided into 4 paths by our demultiplexer, as shown in Fig.\ref{fig:switches} (B). At the starting point, all photons are prepared in horizontal ($\ket{H}$) polarization. When the $1^{th}$ to $25^{th}$ photons arrive at the PC-1-1 and PC-2-1, no voltage signal is applied to the PC. Therefore, the $1^{th}-25^{th}$ $H$ polarized photons will go to spatial mode 1. When the $26^{th}$ to $50^{th}$ $H$ polarized photons arrive at the PC-1-1 and PC-2-1, a $\sim1800$ V half-wave voltage signal is applied to the PC-2-1, the $H$ polarized photons are changed to vertical ($V$) polarized photons. Then a PBS reflects $26^{th}-50^{th}$ $V$ polarized photons to spatial mode 2. If we load all signals as designed in Fig.\ref{fig:switches} (B) to the corresponding PCs as shown in Fig.\ref{fig:switches} (B), we will actively route the $1^{th}-25^{th}$, $26^{th}-50^{th}$, $51^{th}-75^{th}$, $76^{th}-100^{th}$ polarized photons into spatial mode 1, 2, 3, and 4 during each one cycle.

Notice that the time delays among all four output ports are completely different since we only use one single quantum dot in the experiment. To compensate the time delays, single-mode fibers (define well spacial modes) of different lengths are used. We carefully control the lengths of the fibers so that the times of four single-photon streams arrived at the interferometer are nearly identical, which ensure high fiber coupling efficiencies simultaneously. To achieve perfect temporal overlap at the PBSs, every input port of the single-mode fibers is mounted on a translation stage (0.01mm precision and a 25-mm travelling range) are exploited for precisely compensating the time delays for the interference at PBSs. In our experiment, the measured average coupling efficiency is $\sim$85\% (including the loss due to long fibers). Thanks to all anti-reflective optical elements, high coupling efficiency, and high extinction ratio (PCs: extinction ratio $>$100:1; PBSs: extinction ratio $>$2000:1), a measured average single-channel efficiency can be $\sim$83\%.

In the future, one can try to improve fiber end-to-end coupling efficiency to enhance the total efficiency of demultiplexer \cite{alaruri2015singlemodefiber}. Furthermore, one can also try to use a faster optical switch to guide every one photon into different spatial modes. That means, for photonic experiments using few photons, the compensation for time delay and interference between photons from different spatial modes could be directly achieved in free space and the fiber delay will be needless. Then the whole efficiency of the demultiplexer will mainly be limited by the transmission efficiency of optical switches, which principally can be gradually improved to near-unity \cite{wang2017high}.

\hypertarget{part 05}{\subsection{5. A comparison of experimental linear optical CNOT gates}}
\begin{table*}[!h]
	\caption{A comparison of linear optical CNOT gates. SPSs: single-photon sources; PNRDs: photon-number resolving detectors; $P_{s}$: ideal success probability; $\eta_{h}$: heralding efficiency; /: not exit;  --: unreported.}
	\begin{tabular}{|c|c|c|P{2cm}|P{1.5cm}|P{1.5cm}|c|}
		\hline 
		experiments &on-demand SPSs& (pseudo) PNRDs & Heralded &$P_{s}$&$\eta_{h}$&Operation/min\\\hline 
		Pittman \textit{et al.}\cite{pittman2002demonstration}&no&no&no&1/4&/ &$<1$\\ \hline  
		Pittman \textit{et al.}\cite{pittman2003experimental}&no&no&no&1/4&/ &$\sim6$\\ \hline  
		O'Brien \textit{et al.}\cite{o2003demonstration}&no&no&no&1/9&/ &$<1$\\ \hline 
		Kiesel \textit{et al.}\cite{kiesel2005linear}$^{1}$&no&no&no&1/9&/ &$<1$\\ \hline  
		Langford \textit{et al.}\cite{langford2005demonstration}$^{1}$&no&no&no&1/9&/ &$<1$\\ \hline  
		Okamoto \textit{et al.}\cite{okamoto2005demonstration}&no&no&no&1/9&/&$<1$\\ \hline
		Okamoto \textit{et al.}\cite{okamoto2011realization}&no&no&no&1/16&/ &$<2$\\ \hline
		Gasparoni \textit{et al.}\cite{gasparoni2004realization}&no&no&no$^{*}$&1/4&$<10^{-4}$&$<5$\\ \hline  
		Gao \textit{et al.}\cite{gao2010teleportation}&no&no&no$^{*}$&1/9&$<10^{-4}$&$<0.1$ \\ \hline 	  
		Zhao \textit{et al.}\cite{zhao2005experimental}&no&no&no$^{*}$&1/4&$<10^{-6}$&$<1$\\ \hline  
		Bao \textit{et al.}\cite{bao2007optical}&no&no&no$^{*}$&1/8&$<10^{-5}$&$<1$\\ \hline 
		He \textit{et al.}\cite{he2013demand}&yes&no&no&1/9&/ &--\\\hline 
		Gazzano \textit{et al.}\cite{gazzano2013entangling}&yes&no&no&1/9&/ &--\\\hline  
		Pooley \textit{et al.}\cite{pooley2012controlled}$^{1}$&yes&no&no&1/9&/ &--\\\hline   
		Politi \textit{et al.}\cite{politi2008silica}$^{a}$&no&no&no&1/9&/ &--\\\hline  
		Crespi \textit{et al.}\cite{crespi2011integrated}$^{1}$&no&no&no&1/9&/ &--\\\hline 
		Zhang \textit{et al.}\cite{zhang2019femtosecond}$^{1}$&no&no&no&1/9&/&--\\\hline  
		Zeuner \textit{et al.}\cite{zeuner2018integrated}$^{1}$&no&no&no$^{*}$&1/4&$<10^{-4}$&$\sim6$\\\hline  
		Ours &yes&yes&yes&1/8&$\sim$0.008$^{\dagger}$&$\sim$85 \\ \hline 
		\multicolumn{7}{l}{$^{*}$\footnotesize{notes: measurement of output states for postselection due to the multiple pair emissions of SPDC}} \\  
		\multicolumn{7}{l}{$^{\dagger}$\footnotesize{for given input states $\ket{+}_{c}\ket{H}_{t}$; $\eta_{h}$ can be further improved to one in principle.}} \\
		\multicolumn{7}{l}{\footnotesize{$^{1}$ integrated optical implementations}} \\   
	\end{tabular}
	\label{tab:comparisonSI}
\end{table*} 
 
\end{document}